\documentclass[fleqn,usenatbib]{mnras}

\usepackage[T1]{fontenc}

\DeclareRobustCommand{\VAN}[3]{#2}
\let\VANthebibliography\thebibliography
\def\thebibliography{\DeclareRobustCommand{\VAN}[3]{##3}\VANthebibliography}

\usepackage{graphicx}	
\usepackage{adjustbox}
\usepackage{amsmath}	
\usepackage{amssymb}	
\usepackage{xspace}
\usepackage{newtxtext,newtxmath}
\usepackage{comment}

\renewcommand{\vector}[1]{\ensuremath{\pmb{#1}}}
\renewcommand{\div}{\ensuremath{-}}

\newcommand{\Msun}{\ensuremath{\, \rm M_\odot}}
\newcommand{\Msunyr}{\ensuremath{\, {\rm M_\odot}\, {\rm yr}^{-1}}}

\newcommand{\pmag}{\ensuremath{p_{\rm mag}}}

\newcommand{\gcm}{\ensuremath{\,\rm g\, cm^{-2}}}
\newcommand{\gcmc}{\ensuremath{\,\rm g\, cm^{3}}}
\newcommand{\cmsq}{\ensuremath{\,\rm cm^{2}}}
\newcommand{\km}{\ensuremath{\,\rm km}}
\newcommand{\ergl}{\ensuremath{\,\rm erg\,s^{-1}}}
\newcommand{\NS}{\ensuremath{*}}
\newcommand{\xim}{\ensuremath{\xi_{\rm m}}}
\newcommand{\Mach}{\ensuremath{\mathcal{M}}}

\newcommand{\object}[1]{#1} 

\usepackage[normalem]{ulem} 
\def\ver3#1{{#1}}
\defcitealias{BS76}{BS}

\DeclareRobustCommand{\uppartial}{\text{\rotatebox[origin=t]{20}{\scalebox{0.95}[1]{$\partial$}}}\hspace{-1pt}}
\newcommand{\pardir}[2]{\ensuremath{\frac{\uppartial #2}{\uppartial #1} }}

\newcommand{\eint}[2]{\ensuremath{{\rm E}}_{#1}\left(#2\right)}
\renewcommand{\i}{\ensuremath{{\rm i}}}
\newcommand{\diff}{\ensuremath{{\rm d}}}

\title[Accretion column dynamics]{Simulating the shock dynamics of a neutron star accretion column}

\author[Abolmasov \& Lipunova]{
Pavel Abolmasov,$^{1,2}$\thanks{E-mail: pavel.abolmasov@gmail.com}
Galina Lipunova,$^{3,4}$
\\
     $^1$The Raymond and Beverly Sackler School of Physics and Astronomy, Tel Aviv University, Tel Aviv 69978, Israel\\ 
     $^2$ Department of Physics and Astronomy, FI-20014 University of Turku, Finland\\
     $^3$Sternberg Astronomical Institute, Moscow State University,
     Universitetsky pr. 13,  119234 Moscow, Russia\\
     $^{4}$ Max-Planck-Institut f\"ur Radioastronomie, Auf dem H\"ugel 69, 53121 Bonn, Germany \\
}

\date{Accepted XXX. Received YYY; in original form ZZZ}

\pubyear{2022}

\begin{document}
\label{firstpage}
\pagerange{\pageref{firstpage}--\pageref{lastpage}}
\maketitle

\begin{abstract}
   Accretion onto a highly-magnetised neutron star runs through a magnetospheric flow, where the plasma follows the magnetic field lines in the force-free regime.  
  The flow entering the magnetosphere is accelerated by the gravity of the star and then abruptly decelerated in a shock located above the surface of the star. 
  For large enough mass accretion rates, most of the radiation comes from the radiation-pressure-dominated region below the shock, known as accretion column. 
   Though the one-dimensional, stationary structure of this flow has been studied for many years, its global dynamics was hardly ever considered before.
   Considering the time-dependent structure of an accretion column allows us to test the stability of the existing stationary analytic solution, as well as its possible variability modes, and check the validity of its boundary conditions. 
  Using a conservative scheme, we perform one-dimensional time-dependent simulations of an ideal radiative MHD flow inside an aligned dipolar magnetosphere.
   Whenever thermal pressure locally exceeds magnetic pressure, the flow is assumed to lose mass.
  Position of the shock agrees well with the theoretical predictions below a limit likely associated with advection effects: if more than $2/3$ of the released power is advected with the flow, the analytic solution becomes self-inconsistent, and the column starts leaking at a finite height. 
  Depending on the geometry, this breakdown may broaden the column, mass-load the field lines, and produce radiation-driven, mildly relativistic ejecta. 
  Evolving towards the equilibrium position, the shock front experiences damped oscillations at a frequency close to the inverse sound propagation time. 
\end{abstract}




\section{Introduction}

Depending on their nature and fundamental parameters, compact objects may gain matter via different types of accretion flows. 
In particular, for a strongly magnetised neutron star (NS), magnetic field is an important factor shaping the innermost parts of the flow. 
Close to the star, magnetic stresses may dominate the momentum balance, and the matter is forced to follow the field lines and, in particular, to co-rotate with the NS.
Such objects normally manifest themselves as X-ray pulsars (XRPs), hard X-ray sources showing coherent flux pulsations with the spin period of the star \citep[e. g.][]{XRPreview, wolff_baas, Mushtukov-Tsygankov2022}.
For most XRPs, the size of the magnetosphere replacing the disrupted inner portions of the accretion disc is normally two-three orders of magnitude larger than the size of the accreting neutron star (NS). 
Given the shape of the potential well, this means that practically all the energy of the accretion flow is released deep inside the magnetosphere. 
Hence, modelling the magnetospheric part of the flow is crucial for the understanding of the observational properties of XRPs and their long-time evolution. 

XRPs are historically known as a subclass of high-mass X-ray binaries. Most of the well-studied XRPs are binaries consisting of a NS with a massive Be-star companion \citep{bildsten97, raguzova}.
Rapid rotation of the donor stars, large eccentricities and misalignment of many of these systems lead to complicated mass transfer rate variation patterns with different types of flares and orbital and super-orbital variability \citep{reig11}. 
Less numerous but important classes of XRPs include systems with OB-supergiant donor stars (like \object{Vela~X-1}, see \citealt{velaX1} for review) and intermediate-mass subgiant donors (like \object{Her~X-1}, see for example \citealt{herX1}).
Recently, new classes of XRPs were discovered: accreting millisecond X-ray pulsars (see \citealt{AMXP_review} for a review) and super-Eddington pulsating ultraluminous X-ray pulsars \citep{bachetti14}.

Ultraluminous X-ray sources (ULXs) are known since \hbox{1980-s} as a population of extragalactic X-ray sources exceeding the Eddington limit for a conventional black hole (about $10^{39}\ergl$; see \citealt{kaaret} for a review).
They were usually identified as either intermediate-mass black holes in binary systems \citep{PZ04} or stellar-mass black holes accreting at super-Eddington rates \citep{poutanen07}. 
But, intriguingly, during the last few years, several known ULXs were found to produce strong coherent pulsations \citep{bachetti14,fuerst16,israel17a,israel17b,castillo_m51}.
The simplest way to explain the pulsations is by an anisotropically-emitting rotating NS.
Spectral properties of many other ULXs are close to the properties of pulsating ULXs (PULX), suggesting an even larger population of supercritically accreting NSs \citep{pintore17}.
PULX may be considered as XRPs accreting at super-Eddington rates, though a debate is going on about the role of geometric or relativistic beaming in these objects (see, for instance,  \citealt{Abarca+2021} and \citealt{Mushtukov+2021}).

The conventional model of an XRP involves an accretion disc around the NS, a magnetosphere that replaces the inner parts of the disc, and a compact region near the surface of the star where most of the radiation is produced.
Usually, a NS is assumed to have a dipolar magnetic field. However, a pure dipole cannot coexist with the accretion disc outside the outer boundary of the magnetosphere, hence the magnetic field lines are divided into open lines, closed lines (retaining dipole topology), and mass-loaded lines strongly interacting with the disc. 
This picture was proposed in early analytical works \citep[e.g.][]{scharlemann78} and is in general supported by numerical simulations \citep{Kulkarni-Romanova-2013,parfrey17}.
The accretion flow is confined to two narrow flux tubes connecting the inner rim of the disc to the ring- or crescent-shaped regions near the magnetic poles of the NS \citep{bachetti10,Kulkarni-Romanova-2013}. 
Gravitational energy of the matter falling along the magnetic field lines is converted to kinetic energy, and then to heat and to radiation in a shock wave located either very close to the surface of the NS (the hot spot case), or above the surface of the star, if the radiation pressure contribution is large~\citep{Davidson1973, Inoue1975, BS76}.
The latter case, associated with higher luminosities, is usually described as an \emph{accretion column}. 

\ver3{Accretion column structure was considered in a number of papers, analytical and numerical.
An elaborate consideration of the problem has been carried out by \citet{BS76} (hereafter \citetalias{BS76}). 
Below we will consider the `sinking regime' described in section 4.4 of their paper, the case relevant for a geometrically high, radiation-pressure-supported accretion column.
The approximation used by \citetalias{BS76} involved several assumptions, including simplified geometry, stationarity, Newtonian physics, fixed and isotropic scattering cross-section, absence of radial heat diffusion, domination of magnetic stresses over thermal pressure (force-free regime) and radiation pressure over gas one, and a certain set of boundary conditions at the surface of the NS. 
In more recent studies, some of the assumptions of \citetalias{BS76} were relaxed. 
In particular, \citet{arons87,Becker1998,BW07,West+2017} include elaborate microphysics and radiative transfer, while \citet{klein-arons1989, Kawashima+2016,Kawashima-Ohsuga2020,Zhang+2021,zhang22,zhang23} simulate the global time-dependent structure of the flow. 
}

The minimal luminosity required for the formation of a column is less than the Eddington luminosity, as the accretion flow, channelled by the magnetic field lines, is confined to a small fraction of the NS surface.
 The heat in an accretion column is mainly produced by adiabatic heating and radiated away by the sides of the flow. 
The model developed by \citetalias{BS76} recovers the structure of the accretion flow downstream of the shock and determines the position of the shock as an eigenvalue of the problem. 
At the surface of the star, the mass flux is the same as the mass accretion rate at the outer boundary, and the pressure is equal to the magnetic field pressure, meaning the footpoints of the magnetic field lines are at the edge of a breakdown. 
\ver3{Constant mass flux and the presence of a solid NS surface suggest that matter should accumulate inside the column, that is apparently at odds with the stationarity assumption.
Mass accumulation leads to an increase in thermal pressure at the bottom, until the pressure becomes comparable to magnetic stresses, and the force-free approach is no more applicable. 
It is reasonable to assume that this violation leads to a self-regulating mass loss, keeping the column in a steady state. 
The actual physics responsible for this `leakage' is likely some kind of an interchange instability that arises when thermal pressure becomes comparable to magnetic field pressure or exceeds the latter by some factor depending on geometry \citep{litwin01,mukherjee13,Kulsrud-Sunyaev}. }

Here, we aim to relax some of the assumptions of \ver3{the 1D model of} \citetalias{BS76}, \ver3{stationarity in the first place.
As we have seen earlier, the steady-state structure of the column depends upon the process of mass accumulation at the surface and associated pressure growth. 
Thus, time-dependent treatment allows us to check the analytic stationary model of \citetalias{BS76} for consistency.
In particular, we can check if their boundary condition at the surface of the star is a natural outcome of time-dependent evolution.
An important requirement for the simulation is that it should cover a sufficiently long period of time, necessary to accumulate enough mass to fulfil the inner boundary condition.
If the breakdown of the force-free approximation happens at the bottom of the column, the boundary condition and the associated steady-state picture are naturally reproduced. 
However, as magnetic field pressure drops with radius $\propto R^{-6}$, thermal pressure may \hbox{easily} have a shallower profile. 
As we will show in section~\ref{sec:res:breakdown}, this happens in the case of strong advection in the flow. 
If the force-free assumption breaks down above the NS surface, the flow should differ from the steady-state model of \citetalias{BS76}, and may acquire new features such as variability and mass ejections.
}

In the analytic model by \citetalias{BS76}, increase in mass accretion rate shifts the shock wave upwards, towards the accretion disc. 
Applied to PULXs, the model predicts either a shock located very high in the magnetosphere, or a completely subsonic optically thick flow similar to the scenario proposed by \citet{mushtukov17}. 
Extending the solution so far into the magnetosphere requires taking into account additional effects such as irradiation (outer regions of the flow intercept significant part of the luminosity generated by its inner parts), centrifugal force, and more complex geometry (dipole is no more consistent with a power-law approximation).
Apart from this, as we will see in section~\ref{sec:res:compare}, self-consistency of the analytic model is limited by advection effects that become important for large mass accretion rates and small radiating surfaces of the column.

An important question in the case of NSs is whether accretion columns have observable global oscillation modes. 
This is potentially a very important issue, because any oscillations formed near the surface of the star are direct probes of its properties: strong gravity, magnetic fields, and the physical conditions in the flow. 

Apparently, there is no unambiguous observational evidence for any oscillation modes coming from accretion columns.
Normally, the power density spectrum (PDS) of an XRP shows a broad-band noise with a break at some frequency positively correlated with the flux \citep{revnivtsev09}. 
The break frequency is close to the expected Keplerian frequency at the boundary between the magnetosphere and the disc, and the observed correlation reasonably fits into the concept of the magnetosphere size changing  with the accretion rate~\citep[see, e.g.,][]{pringle-rees1972,Lamb_etal1973,bildsten97,Filippova+2017}.
However, at large, super-Eddington, luminosities, the PDS acquires an additional broad-band noise component peaking at about the break frequency \citep{revnivtsev09, RN13}. 
One of the proposed explanations is the presence of oscillations in the outer parts of the magnetospheric flow. 
The possible sources of the oscillations could be the feedback from the irradiation by the inner, bright parts of the flow, or a combination of the centrifugal force, radiation, and gravity.
One such solution involving low-optical-depth matter trapped in the magnetosphere was proposed by \citet{AB20}.

At higher frequencies (closer to the dynamic frequencies near the surface of the star) most XRPs lack detectable variability. 
There are however two notable exceptions which will be discussed in this paper (see section~\ref{sec:disc:osc}).
One is GRO~J1744$-$28 \citep{klein96}, where a significant power excess at tens of Hz is observed.
The frequency range of the noise is suggestive of the dynamical time scale of the accretion column in this object.
Another example is \object{Cen~X-3}, for which \citet{jernigan00} detected a broad, low-quality-factor feature at about 1\,kHz (see, however, \citealt{revnivtsev15}). 
Both cases were explained by the authors as manifestations of the photon-bubble instability. 
In this paper, we show that similar observational features may be reproduced by global oscillations of the accretion column.

The structure of the paper is as follows.
We formulate the problem, write down the equations, and describe the numerical code in section~\ref{sec:problem}. In section~\ref{sec:res}, we present the results including velocity profiles, shock positions, and variability patterns. We discuss the results in section~\ref{sec:disc}, and conclude in section~\ref{sec:conc}.

\section{Problem formulation and numerical setup}\label{sec:problem}

We consider the dynamics of a flow restricted by the aligned dipolar magnetic field of the NS. 
Motion along the magnetic field lines is affected by  gravity, pressure gradients, centrifugal force, and radiation pressure.
We use one-dimensional time-dependent approach, assuming the accretion flow uniformly fills a flux tube. 
Within the tube, all the physical parameters depend only on time and  the coordinate along the field line.
We use the laws of conservation of mass (continuity equation), momentum (Euler equation projected along the field line), and energy. 
The code is freely available at \url{https://github.com/pabolmasov/HACol}.

\begin{figure*}
\adjincludegraphics[width=0.75\textwidth,trim={0cm 3.5cm 0cm 1cm},clip]{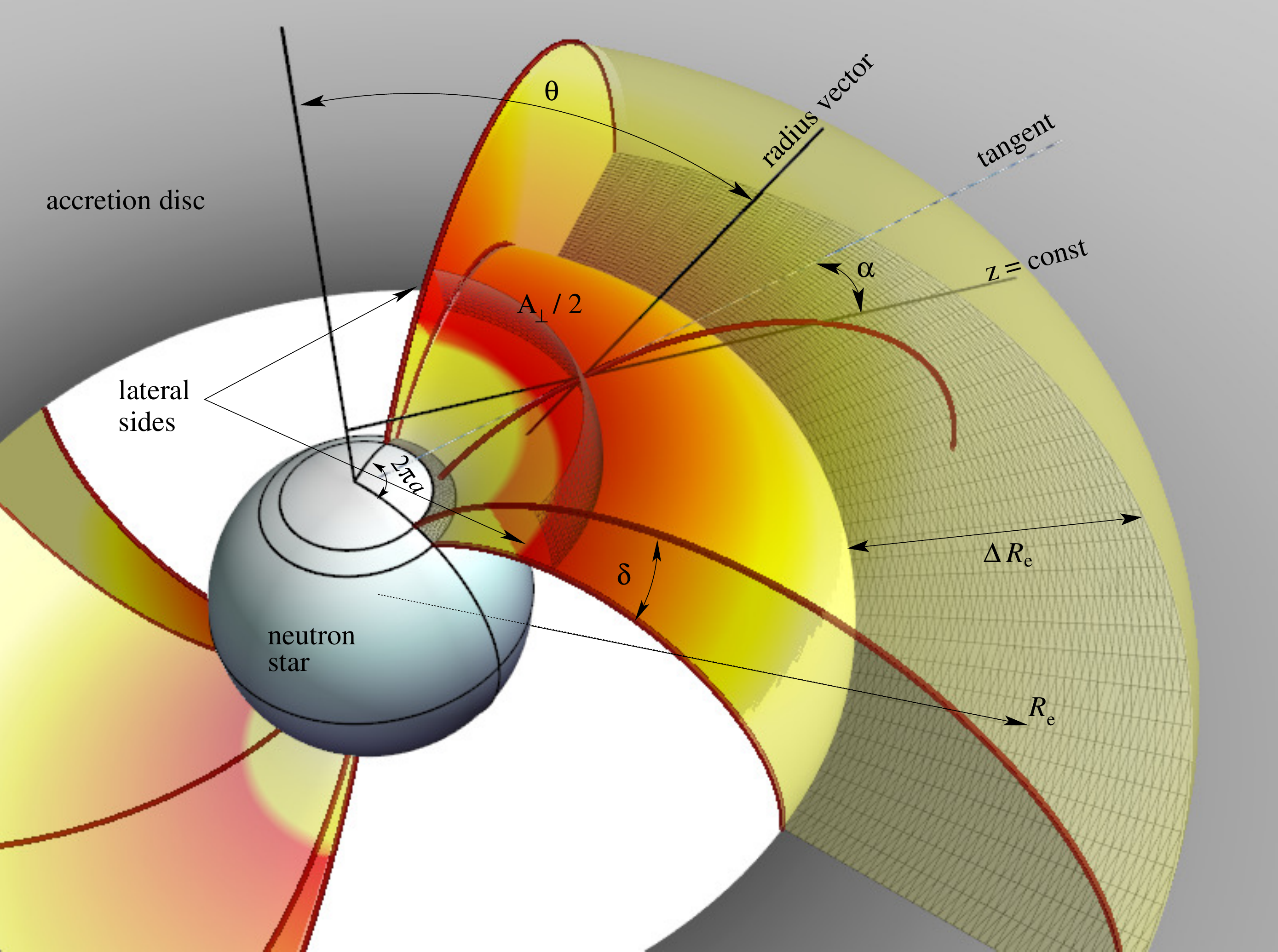}
 \caption{A sketch illustrating the geometry of the problem. 
 Red solid lines are individual magnetic field lines: four on each side restricting the flow and one in the middle. 
 Three cross-sections are shown with hatched shadings, one on the surface, one above the surface, and one in the equatorial plane (outer boundary).  
 The blue dashed line shows a tangent to the field line at the distance where the second cross-section is plotted. 
 We also show radial and $z=$const lines passing through the same point. 
 Colouring of the sides of the flow shows the pressure (normalised by local magnetic pressure) distribution in the flow according to one of the simulations: yellow corresponds to values about one, red is about $10^{-3}$. 
 The colouring in this plot is schematic and does not correspond to any particular model.
 }\label{fig:nssketch}       
\end{figure*}

\subsection{Geometry}

Figure~\ref{fig:nssketch} shows the adopted geometry of the problem. 
The simulated flow is a pair of symmetric flux tubes, restricted by the surfaces $\theta = \mbox{const}$ and $\varphi = \mbox{const}$, starting in the accretion disc and ending on the surface of the NS. 
The field lines are the lines of an unperturbed magnetic dipole.
We use a spherical coordinate grid: radius $R$, polar angle $\theta$, and azimuthal angle $\varphi$. 
For a single field line, radial coordinate $R = R_{\rm e} \sin^2\theta $, where $R_{\rm e}$  is the equatorial size of the magnetosphere. 
 By the order of magnitude, the size of the magnetosphere is equal to the Alfv\'{e}n radius
\begin{equation}\label{E:geo:Alfven}
    R_{\rm A} = \left( \frac{\mu^2}{2\dot{M}\sqrt{2GM_\NS}}\right)^{2/7}.
\end{equation}
Here, $\dot{M}$ is the mass accretion rate, $\mu$ is the magnetic moment of the NS, $M_\NS$ is the mass of the NS.
We normalise the size of the magnetosphere as $R_{\rm e} = \xim R_{\rm A}$ and set $\xim = 0.5$ in all our models.

For two field lines having slightly different equatorial radii close to $R_{\rm e}$ and separated at the equator by $\Delta R_{\rm e} \ll R_{\rm e}$, the distance between the field lines $\delta$ is expressed through the polar angle as (see Appendix~\ref{app:geo}):
\begin{equation}\label{E:geo:delta}
\delta = \frac{R \sin\theta}{\sqrt{1+3\cos^2\theta}} \frac{\Delta R_{\rm e}}{R_{\rm e}} .
\end{equation}
Here, $\Delta R_{\rm e}$ has the physical meaning of the penetration depth of the magnetic field into the accretion disc (see Fig.~\ref{fig:nssketch}).
Relatively small value of $\Delta R_{\rm e}$ is a consequence of the high conductivity of the plasma and is reproduced by numerical simulations \citep{parfrey16, romanova03}.

A thin (in the sense of $\delta \ll R$) flow has a cross-section of 
\begin{equation}\label{E:geo:across}
\displaystyle  A_\perp =  2\, \delta\, 2\uppi a R\sin\theta = 4\uppi a R_{\rm e} \Delta R_{\rm e} \frac{\sin^6\theta}{\sqrt{1+3\cos^2\theta}}\, .
\end{equation}
We assume that in the azimuthal direction, the flow occupies a fixed fraction $a$ of the full circle $2\uppi R\sin\theta$, with $0<a\leq 1$. 
In the absence of strict axisymmetry, the flow is expected to occupy only a limited range of azimuthal coordinates.
We treat the cross-section as rectangular, with the perimeter equal to
\begin{equation}\label{E:geo:perimeter}
   \Pi =  2 \, \left(  \frac{A_\perp}{\delta} + 2\delta\right),
\end{equation}
where the first factor of 2 (as well as the first multiplier $2$ in equation~\ref{E:geo:across}) comes from the flow actually consisting of two streams accreting onto the two poles of the dipole. 
Taking into account the lateral sides of the flow (the second term in equation~\ref{E:geo:perimeter}) increases the surface of the column by a factor $1+2\delta^2/A_\perp$.

As an independent variable for calculations, it is convenient to use a coordinate along the field line, defined as (see equation~\ref{E:cosa}):
\begin{equation}\label{E:geo:l}
 \displaystyle  l = \int \frac{\sqrt{3\cos^2\theta+1}}{2\cos\theta}\, {\rm d}R = R_{\rm e}\int \sqrt{3\cos^2\theta+1} \sin\theta \,{\rm d}\theta\, .
\end{equation}
As both variables, $R$ and $\theta$, increase monotonically along the field line, conversion between $l$, $R$, and $\theta$ is unique and straightforward. 

\subsection{Time scales}\label{sec:times}

For each radius $R$, there is a characteristic dynamical time
\begin{equation}\label{E:time:dyn}
    t_{\rm d}(R) = 2\uppi \sqrt{\frac{R^3}{GM_\NS}}. 
\end{equation}
At the surface of the NS, it is of the order $t_{\rm d}(R_{\NS}) \sim 10^{-4}$s. 
At the outer edge of the magnetosphere, it can reach the values of seconds, or even larger, for strong enough magnetic fields or small mass accretion rates.

In section~\ref{sec:res:var} we will also discuss the sound propagation time scale $t_{\rm s}$ along the field line
\begin{equation}\label{E:time:sound}
    t_{\rm s} = \int \frac{{\rm d}l}{c_{\rm s}},
\end{equation}
where $c_{\rm s}$ is the speed of sound.
This time depends on the structure of the column, and may be longer (because of sub-virial speed of sound) as well as shorter (when the height of the column is small compared to $R_*$) than local dynamical times. 

Another important time scale is the \emph{replenishment time} of the matter in the column, $t_{\rm r} = M_{\rm col} / \dot{M}$, where 
\begin{equation}
    M_{\rm col} = \int_{R_{\NS}}^{R_{\rm shock}} A_\perp \rho \,{\rm d}R
\end{equation}
is the mass of the accretion column (of all the matter below the shock and above the surface of the star). 
For a crude estimate, let us assume that $R_{\rm shock} - R_{\NS} \ll R_{\NS}$, and the column itself is in hydrostatic equilibrium. This allows us to replace $\rho {\rm d}R = - {\rm d}p/g$ (where $g=GM_\NS/R_\NS^2$ is the gravity at the NS surface), hence
\begin{equation}\label{E:mcol}
    M_{\rm col} \simeq \left(A_\perp p  / g\right)_{R = R_{\NS}}.
\end{equation}
Applying the lower boundary condition $p = B^2/8\uppi$ results in a column mass estimate independent of the mass accretion rate. 
Making all the necessary substitutions, we get an estimate for the replenishment time scale
\begin{equation}\label{E:trep}
\begin{array}{l}
    \displaystyle t_{\rm r} \simeq \frac{B^2}{8\uppi GM} \frac{A_\perp(R_*) R_*^{2}}{\dot M}\\ 
   \displaystyle \qquad{}  \simeq 340 \, \left( \frac{B}{10^{12}\,\rm G} \right)^2 \frac{A_\perp(R_{\NS})}{10^{12}\cmsq} \left( \frac{R_\NS}{10\km }\right)^2 \frac{10^{-8}\Msunyr}{\dot{M}}\,\mbox{s}, 
   \end{array}
\end{equation}
For pulsar-scale magnetic fields and Eddington mass accretion rate $\dot{M} \simeq 10^{-8}\Msunyr$, $A_\perp(R_{\NS}) \sim 0.1 R_{\NS}^2$ (see equation~\ref{eq.Aperp}), and $t_{\rm r}$ is of the order of seconds or tens of seconds. 
The magnetospheres considered in this paper are generally smaller and mostly have $t_{\rm r} \sim 0.1$s, depending on the geometric parameters such as $a$ and $\Delta R_{\rm e}$. 

Replenishment time scale is evidently longer than the dynamical time at the surface of the star. 
It is also likely to exceed the dynamical time scale at the outer edge of the magnetosphere. 
Using equation~(\ref{E:geo:Alfven}), it is possible to write down the ratio of the two times as
\begin{equation}\label{E:trep:RA}
\frac{t_{\rm r}}{t_{\rm d}(R_{\rm e})} = \frac{a\sqrt{2}}{\uppi\,\xim^{7/2}} \frac{\Delta R_{\rm e}}{R_{\NS}}.
\end{equation}
As $\xim \sim 1$, $\Delta R_{\rm e} \lesssim R_{\rm e}$, and $R_\NS \ll R_{\rm e}$, the replenishment time is generally longer than the dynamical time scale at the size of the magnetosphere.

Another crucial time, the time scale for thermal cooling, may be defined as the time required to lose the local internal energy by radiation. 
In the optically thick parts of the flow, and given $\delta^2 \ll A_\perp$, it may be calculated as the radiation diffusion time scale $\delta^2/D$, where $D = c/3\varkappa \rho$ is the radiation diffusion coefficient. 
For a stationary flow, 
\begin{equation}\label{E:tthermal}
    t_{\rm thermal} \sim \frac{3\varkappa \dot{M}}{cv} \frac{\delta^2}{A_\perp},
\end{equation}
where $\varkappa$ is the opacity. 

Velocity $v$ varies along the field line from the free-fall velocity above the shock wave to seven times slower below the shock and then essentially to zero near the surface. 
This makes thermal time scales extremely long near the basement of the column. 
To estimate the characteristic maximal cooling times, we can use the analytic
solution of \citetalias{BS76}.
The velocity may be obtained by dividing the two equations (32) of \citetalias{BS76} one over the other and substituting $R=R_\NS$, that yields for the velocity at the basement of the column
\begin{equation}
    v(R_\NS) \simeq \frac{2\uppi \beta_{\rm BS}}{B^2} \frac{GM\dot{M}}{R_\NS A_\perp(R_\NS)}\, ,    
\end{equation}
where $\beta_{\rm BS}$ is the `$\beta$' parameter used by \citetalias{BS76}, that has the physical meaning of the fraction of accretion power advected through the surface of the star. 
For realistic parameter sets, $\beta_{\rm BS}$ varies approximately between $0.1$ and $1$. 
Substituting this velocity estimate to (\ref{E:tthermal}), we get an expression for the thermal time scale at the bottom of a 
column
\begin{equation}\label{E:tthermal:bottom}
\begin{array}{l}
   \displaystyle t_{\rm thermal}(R_\NS) \simeq \frac{3B^2}{2\uppi \beta_{\rm BS}} \frac{\varkappa}{GMc} R_\NS \delta^2(R_\NS)\\
    \displaystyle  \qquad{} \simeq 300 \frac{1}{\beta_{\rm BS}} \left( \frac{B}{10^{12}{\rm G}}\right)^2 \left( \frac{\delta(R_\NS)}{1\,\rm  km} \right)^2 {\rm s}\, .\\
\end{array}
\end{equation}
The bottom of the column is interacting with the atmosphere, ocean, or solid surface of the NS. 
The pressure may be estimated as $p_{\rm bottom} \sim B^2/8\uppi$. 
Equivalent column density $y = p_{\rm bottom} / g \simeq 2\times 10^8 \left( B/10^{12}\rm G\right)^{2} \gcm$ likely corresponds to the ocean layer of a cold NS surface. 
Following \citet[section 7.1]{meisel18}, one can estimate the thermal time scale within the ocean or cold atmosphere as $t_{\rm ocean} \simeq 100  ~{\rm s}\,\left( B/10^{12}{\rm G}\right)^{3/2}$.
\ver3{For typical parameter values (see Table~\ref{tab:mod})}, this estimate is \ver3{few times longer than} the thermal time scale in the lower parts of the column (Eq.~\ref{E:tthermal:bottom}) and significantly longer than the replenishment time (\ref{E:trep}).
Hence, thermal interaction with the NS surface is likely a very slow process, and accretion column may be considered thermally isolated from the NS.

\subsection{Conservation equations}\label{sec:num:con}

For a one-dimensional formulation, the basic density quantities (mass density, momentum density, and energy density) need to be integrated in the direction perpendicular to the magnetic field lines. 
Assuming the physical conditions do not significantly vary across the flux tube, we replace this integration with multiplication by the cross-section $A_\perp $ given by equation~(\ref{E:geo:across}). 
Computation involves a finite-volume conservative scheme for the three quantities we treat as conserved: mass, momentum along the field line, and energy, expressed per unit length $l$ along the flux tube (from the computational point of view, it is a single field line)
\begin{equation}\label{E:con:m}
  m = \int \rho\, {\rm d}A =  \rho A_\perp, 
\end{equation}
\begin{equation}\label{E:con:s}
  s = \int \rho v\, {\rm d}A =  \rho v A_\perp , 
\end{equation}
\begin{equation}\label{E:con:e}
\displaystyle  e = \int \left(u + \rho\, \frac{v^2}{2}\right) {\rm d}A = \left(u + \frac{\rho\, v^2}{2}\right) A_\perp. 
\end{equation}
Here, $\rho$ is the volume rest-mass density, $v$ is the velocity along the field line,  $u$ is the thermal energy density, consisting of gas $u_{\rm gas}$ and radiation $u_{\rm rad}$ energy densities.
We use a kinematic constraint that the matter does not move across the magnetic  field lines, and the field lines themselves are not distorted. 
This justifies the use of a rotating frame and the absence of the kinetic energy of rotation in (\ref{E:con:e}).
For each of the three quantities, conservation laws have the general form
\begin{equation}\label{E:con:qcons}
  \pardir{t}{q} + \pardir{l}{F_q} = S_q,
\end{equation}
where $q$ refers to a particular quantity ($m$, $s$, or $e$), and $F_q$ and $S_q$ are, respectively, the corresponding flux and the source term. 
We take the fluxes in the form
\begin{equation}\label{E:flux:m}
\displaystyle F_m =  \int \rho v \,{\rm d}A = s,
\end{equation}
\begin{equation}\label{E:flux:s}
\displaystyle F_s =  \int \left(\rho v^2 + p\right) \,{\rm d}A = s v + A_\perp p,
\end{equation}
and
\begin{equation}\label{E:flux:e}
\begin{array}{l}
\displaystyle F_e =  \int \left( v \rho \left(\frac{u + p}{\rho} + \frac{v^2}{2} \right)- D \pardir{l}{u_{\rm rad}}\right) \,{\rm d}A  \\
\qquad{} \displaystyle = s \left(\frac{u + p}{\rho} + \frac{v^2}{2} \right) - D A_\perp \left(\pardir{l}{u_{\rm rad}}\right) ,\\
\end{array}
\end{equation}
where $D = c/3\varkappa \rho$ is the radiation diffusion coefficient introduced in Section~\ref{sec:times}.
The second term of equation~(\ref{E:flux:e}) accounts for photon diffusion along the field lines.
Thermal pressure $p$ used in equation~(\ref{E:flux:e}) includes gas and radiation pressure, and is related to the internal energy density as
\begin{equation}\label{E:pressure}
  \displaystyle   p = \frac{u}{3 \left( 1-\beta /2\right)},
\end{equation}
and $\beta$ is defined as the ratio of gas pressure to the total.  
Pressure ratio may be found by solving numerically (for derivation see Appendix~\ref{app:beta}) the equation for $\beta$
\begin{equation}\label{E:beta}
\displaystyle 
\frac{\beta}{\left(1-\beta/2\right)^{3/4}\left( 1-\beta\right)^{1/4}}  = \frac{3}{\sqrt{2}} \frac{k}{\tilde{m}} \left(\frac{c}{\sigma_{\rm SB}} \right)^{1/4} \frac{\rho}{u^{3/4}}\, .
\end{equation}
Here, $k$ and $\sigma_{\rm SB}$ are, respectively, the Boltzmann and
Stefan-Boltzmann constants, and $\tilde{m}$ is the mean particle mass that we set to $\tilde{m}= 0.6\, m_{\rm p}$. 

As the accreted matter tends to accumulate at the surface of the star, to reach a stationary state, we need a physically motivated mass sink. 
We add a mass sink proportional to the density and perimeter that turns on when the local thermal pressure exceeds the pressure of the field $\pmag =B^2/8\uppi$, 
\begin{equation}\label{E:src:m}
    S_m = - \frac{m}{A_\perp}\, \Pi\, \sqrt{\frac{\Gamma_1 \max\left(p-\pmag, \ 0\right) }{\rho}}.
\end{equation}
Here, $\Pi$ is the perimeter of the flow given by equation~(\ref{E:geo:perimeter}), and $\Gamma_1$ is the effective adiabatic exponent for ideal monatomic gas with a contribution of radiation pressure, see \citet[Chapter II, equation 131]{chandra67} 
\begin{equation}\label{E:gamma1}
 \displaystyle   \Gamma_{1} = \beta + \frac{\left(4-3\beta\right)^2\left(\Gamma-1\right)}{\beta + 12 \left(\Gamma-1\right) \left( 1-\beta\right)},
\end{equation}
and $\beta =p_{\rm gas}/p$ is calculated using equation~(\ref{E:beta}), and $\Gamma=5/3$ as for ideal monatomic gas. 
Magnetic field on the surface is calculated according to the dipole formula (see Appendix~\ref{app:geo}).

The loss term given by equation~(\ref{E:src:m}) is zero when thermal pressure is low, turns on when $p=p_{\rm mag}$, and grows continuously with increasing pressure. 
For large pressure, the loss term corresponds to all the excess matter escaping the flux tube at about sonic velocity.
For the innermost cell of the simulation, it also works as a boundary condition ensuring that, for a column mass high enough, the energy density at the bottom of the column conforms with the boundary condition used by \citetalias{BS76}. 

\ver3{
The sideways motion of the plasma is expected due to pressure-driven instabilities considered by 
\citet{litwin01, mukherjee13,Kulsrud-Sunyaev} for the matter at the NS surface.
One-dimensional approach does not allow us to simulate transverse motions, hence we include them as sink terms in conservation equations. 
Growth rates $\sim  10^5$~s$^{-1}$ estimated by \citet{Kulsrud-Sunyaev} indicate that the characteristic times are well below the dynamical time justifying our assumption that the leak instantly adjusts to the current physical state of the column.
We also neglect the associated contribution to kinetic energy density $\rho v_\perp^2/2$ (where $v_\perp$ is transverse velocity), assuming the bulk of the column is not involved in the spreading motion.}

\ver3{Higher up the column, away from the fixed footpoints of the field lines, Kruskal-Schwarzschild-type instability may be responsible for the mass loss, if thermal pressure is comparable to the local magnetic field pressure.
Applying Bernstein's energy principle \citep{Bernstein+1958} to the column surface and assuming the column is in hydrostatic balance, one obtains a high growth rate $\gtrsim v_{\rm A}/\sqrt{R\delta }$~\citep[see eq.~5.9,][]{Bernstein+1958}, where $v_{\rm A}$ is the Alfv\'en velocity, which apparently means that leaks are not time-resolved in our setting.}

The source term for the momentum $s$ is
\begin{equation}\label{E:src:s}
S_s = g_\parallel m + S_m v,
\end{equation}
where $g_\parallel$ is the force (gravitational, centrifugal, and radiation pressure) per unit mass acting onto the matter in the flow, 
\begin{equation}\label{E:g_parallel}
    g_\parallel = - \frac{GM_*}{R^2} \sin(\alpha + \theta) \left( 1-\Gamma_{\rm Edd}\right) + \Omega^2 R \sin\theta \cos\alpha,
\end{equation}
$\alpha$ is the angle between the tangent to the field line and the direction of the centrifugal force (see Fig.~\ref{fig:nssketch}),
$\Omega$ is the rotation frequency of the NS, and
\begin{equation}\label{E:eddfactor}
    \Gamma_{\rm Edd}=\eta_{\rm irr} \frac{L}{L_{\rm Edd}} \frac{1-{\rm e}^{-\tau}}{\tau}
\end{equation}
is the correction for the radiation pressure created by the irradiation of the flow, i.e. by the photons that come to a particular point  from any other point: the surface of the neutron star, the surface of the second accretion column, etc. 
We assume that $\eta_{\rm irr}\lesssim 1$ and constant,  $L$ is the total power lost by the flow as radiation, the Eddington luminosity 
\begin{equation}
    \displaystyle L_{\rm Edd} = \frac{4\pi G M_\NS c}{\varkappa} \simeq 2\times 10^{38} \frac{M_\NS}{1.4\Msun}\ergl,
\end{equation}
and $\tau$ is the optical depth
across the flow in the poloidal direction, estimated as 
\begin{equation}\label{E:tau}
\displaystyle \tau \simeq \tau_\theta = \varkappa \rho \delta = \varkappa  \frac{\delta}{A_\perp} m \,  .
\end{equation}
The factor $\eta_{\rm irr}$ accounts for our poor knowledge about the geometry and anisotropy of the radiation field. 
The irradiation can play very important role for 
\ver3{the flow above the shock, where the optical depth is relatively low,} but it is suppressed by a factor of $\tau \gtrsim 100$ in the column itself.
The probability of a photon emitted by some part of the column to intersect a mass-loaded field line is large if $a \sim 1$.

The last term in (\ref{E:src:s}) accounts for the momentum lost with the expelled mass. 

In this work, we use the opacity of $\varkappa = 0.35\, \rm cm^2\, g^{-1}$, approximately equal to the Thomson scattering opacity for Solar metallicity in the assumption of complete ionisation. 
We neglect the influence of the magnetic field upon the scattering cross-section, that could be an important factor for large magnetic fields \citep{BS75,arons87,BW07}.

Here, we consider three different contributions to energy sources and sinks: work done by the external forces, energy loss due to radiation, and the energy lost with the mass loss
\begin{equation}\label{E:src:e}
 \displaystyle  S_e = g_\parallel s - \frac{1-{\rm e}^{-\tau_{\rm eff}}}{\xi_{\rm rad}\tau_{\rm eff}+1} \,c\,  \Pi u_{\rm rad} 
 + S_m \frac{e+p A_\perp}{m},
\end{equation}
where $\xi_{\rm rad}\sim 1$ is a dimensionless factor taking into account the transverse structure of the flow.
When energy is released in the middle of the accretion column, $\xi_{\rm rad} = 3/2$. 
Assuming the energy sources uniformly distributed over the cross-section would instead lead to $\xi_{\rm rad} = 3/4$. 
Both possibilities are mentioned in \citetalias{BS76}, and the first one used for the calculations.
We also use $\xi_{\rm rad} = 3/2$ everywhere in this paper. 
The $1-e^{-\tau_{\rm eff}}$ factor in the second term of equation~(\ref{E:src:e}) allows us to extend the results to the case of low optical depths, where the radiation losses are volumetric rather than areal.
We use the effective optical depth defined as
\begin{equation}\label{E:taueff}
    \tau_{\rm eff} = \varkappa\, m\, \delta_{\rm eff}/A_\perp,
\end{equation}
where
\begin{equation}
\delta_{\rm eff}^{-1} = \delta^{-1} + 2\delta/A_\perp .
\end{equation}
For the models where emission from the lateral sides is ignored, $\delta_{\rm eff} = \delta$, $\tau_{\rm eff}= \tau_\theta$, and perimeter $\Pi = 2A_\perp / \delta$.
equation~(\ref{E:taueff}) reproduces the optical depth in the two extreme cases, when $A_\perp / 2 \delta$ is either much smaller or much larger than $\delta$.
The last term in~(\ref{E:src:e}) corresponds to the matter lost from the column when the mass sink is on.
As work should be done to expel matter, there is a contribution from pressure in the term. 
A more accurate expression for the term should account for the transverse structure of the column and multiple energy transformation processes happening during the breakdown of the force-free regime.

The resulting system of three differential equations takes the form 
\begin{equation}\label{E:system:mass}
    \pardir{t}{m} + \pardir{l}{s} = S_m,
\end{equation}
\begin{equation}\label{E:system:momentum}
    \pardir{t}{s} + \pardir{l}{F_s} = S_s,
\end{equation}
and
\begin{equation}\label{E:system:energy}
    \pardir{t}{e} + \pardir{l}{F_e} = S_e,
\end{equation}
where the densities, fluxes, and sources are given by equations (\ref{E:con:m}-\ref{E:con:e}), (\ref{E:flux:m}-\ref{E:flux:e}), (\ref{E:src:m}), (\ref{E:src:s}), and (\ref{E:src:e}). 
The independent variable $l$ is related to the radial coordinate by equation~(\ref{E:geo:l}). 
The cross-section of the flow used by the expressions for fluxes and sources is calculated according to equation~(\ref{E:geo:across}).

Solving the above system of equations yields physical parameters as functions of time $t$ and spherical radius $R$. 
The bolometric luminosity is calculated by integration of  the second term of \eqref{E:src:e} along the flux tube as
\begin{equation}\label{E:ltot}
L = \int Q^- \Pi\,{\rm d}l = \int \frac{1-{\rm e}^{-\tau_{\rm eff}}}{\xi_{\rm rad}\tau_{\rm eff}+1} \,c\,  \Pi\, u_{\rm rad}
\,{\rm d}l,
\end{equation}
where perimeter $\Pi$ is given by equation~(\ref{E:geo:perimeter}), $\tau_{\rm eff}$ by equation~(\ref{E:taueff}), and 
\begin{equation}\label{E:qminus}
    Q^- = \frac{1-{\rm e}^{-\tau_{\rm eff}}}{\xi_{\rm rad}\tau_{\rm eff}+1} \,c\,  u_{\rm rad}
\end{equation}
is the radiation flux leaving the surface of the column.
The integral in equation~(\ref{E:ltot}) taken over the whole simulation domain will be denoted $L_{\rm tot}$, and below the shock $L_{\rm X}$, for consistency with \citetalias{BS76}.

\subsection{Numerical method}

System of equations (\ref{E:system:mass}--\ref{E:system:energy}) is solved using HLLE (Harten -- Lax -- van Leer -- Einfeldt) Riemann solver \citep{HLL, HLLE_Einfeldt}.
 Signal velocities used in this solver are calculated according to the `hybrid' method of \citet{toro94}, with the effective adiabate exponent set to $\Gamma_{\rm eff} = 5/3$. Hence, the adopted maximal signal velocity is slightly larger than the accurate value predicted by equation~(\ref{E:gamma1}). 
This adds diffusivity to the solution but increases its stability in the case of rapid variations of $\Gamma_1$.

The problem is challenging for numerical consideration, as density and velocity vary by many orders of magnitude. 
For the configurations considered in this paper, Mach number \Mach\ changes from tens above the shock front to about $10^{-5}$ near the surface of the NS. 
While existence of a shock front is not a problem for an HLLE solver, low Mach values cause the so-called stiffness problem \citep{lowmach}. 
In eliminating the problem, we find the pre-conditioning technique introduced by \citet{turkel99} sufficient for our needs. 
For Mach numbers smaller than $1$, signal velocities used by the Riemann solver are multiplied by \Mach, that efficiently removes the spurious flux biases in the low-velocity regions. 

The purpose of a Riemann solver is conversion between the fluxes calculated at the midpoints of the cells and the fluxes at the cell boundaries. 
The energy flux given by equation~(\ref{E:flux:e}) contains a photon diffusion term that is immediately calculable at the cell boundaries as
\begin{equation}
    -D A_\perp \pardir{l}{u_{\rm rad}} \simeq \frac{c A_\perp}{3} \frac{u_{{\rm rad,} \, i+1} - u_{{\rm rad,} \, i}}{\tau_i} 
\end{equation}
where $\tau_i = \varkappa \int_{l_{i}}^{l_{i+1}} \rho \,{\rm d}l \simeq (\rho_i + \rho_{i+1}) (l_{i+1}-l_i)/2$ is the optical depth between the centers of the $i$-th and the $i+1$-th cells. 
The diffusive energy flux term is limited in its absolute magnitude by the maximal possible value, $c A_\perp \min(u_{{\rm rad}, i}, u_{{\rm rad}, i+1})$.
The standard HLLE solver is used for the rest of the flux terms. 

\subsection{Boundary and initial conditions}\label{sec:num:BC}

As the inner boundary conditions, we use zero velocity (hence $F_m(R_\NS) = s(R_\NS) = 0$) and zero thermal energy flux (that implies $F_e(R=R_*) = 0$). 
In the lower parts of the column, mass and energy accumulate until the pressure of the flow exceeds the local magnetic field pressure, and the mass-loss term (\ref{E:src:m}) turns on. 
If the mass loss starts at the bottom, the boundary condition used by \citetalias{BS76} is reproduced. 

At the outer boundary, we fix the mass flux $F_m$ (and thus momentum density $s=F_m$) to $-\dot{M}$.
The material entering at the outer boundary is assumed to have the velocity of $v(R_{\rm e}) = -\sqrt{GM_\NS/R_{\rm e}}$.
Knowing the mass flux and the velocity at the outer boundary allows one to calculate the value of density $\rho$.
Thermal energy density $u$ at the outer boundary was chosen equal to the magnetic energy density at this distance.
As the transverse optical depth of the flow above the shock is usually small, most of this internal energy is radiated away above the shock wave.
Hence, the flow is always gravitationally bound. 

The total energy radiated by a steady-state flow is
\begin{equation}
\label{E:energy_conservation}
    L_{\rm tot} = \frac{GM_\NS\dot{M}}{R_\NS}-\frac{GM_\NS\dot{M}}{2R_{\rm e}} + L_{\rm out} - L_{\rm vent}\, ,
\end{equation}
where 
\begin{equation}\label{E:lout}
   \displaystyle L_{\rm out} =\dot{M}  \left. \frac{u+p}{\rho}\right|_{R_{\rm e}}
\end{equation}
is the heat entering through the outer boundary per unit time, and $L_{\rm vent}$ is the energy loss associated with the mass loss from the column (third term of equation~\ref{E:src:e} integrated along the field line).
The quantities $\dot{M}$, $\rho$, and $u$ in equation~(\ref{E:lout}) are set by the boundary conditions, and $p$ may be derived from $\rho$ and $u$ using the formulae of Appendix~\ref{app:beta}.
Usually the first term of equation~(\ref{E:energy_conservation}),
\begin{equation}
    \displaystyle L_{\rm acc} = \frac{GM\dot{M}}{R_*},
\end{equation}
dominates, and the thermal contribution of the outer boundary condition is small. 
However, for low accretion rates, the latter can significantly alter the total  luminosity of the flow.
On the other hand, the measured steady-state luminosity may be (and usually is) smaller than $L_{\rm acc}$ because of the adopted mass and the energy sink represented by $L_{\rm vent}$. 

Initial conditions are approximately constant mass density and sub-virial negative velocity approaching zero at the inner boundary. 
We choose the total initial mass equal to 10 per cent of the equilibrium column mass defined as (equation~\ref{E:mcol}), that allows the flow to approach a quasi-stationary regime in several replenishment times (equation~\ref{E:trep}). 
The relatively slow approach to equilibrium (see section~\ref{sec:res}) is probably related to very long thermal time scale at the bottom of the column (equation \ref{E:tthermal:bottom}).

\subsection{Time steps}

Ignoring the effects of radiation losses, we can estimate the time step required for stability as
\begin{equation}\label{E:dt:CFL}
    \Delta t_{\rm CFL} \simeq C_{\rm CFL} \left(\frac{\Delta l}{|v|+c_{\rm s}}\right)_{\rm min},
\end{equation}
where $C_{\rm CFL} \lesssim 1$ is the Courant-Friedrichs-Levy multiplier \citep{CFL}. 
However, local thermal radiation loss and diffusion time scales sometimes become smaller than  \eqref{E:dt:CFL},  and this
requires refinement of the time step.
Thermal cooling time scale may be estimated using the radiation loss term in equation~(\ref{E:src:e}), as
\begin{equation}\label{E:dt:tthermal}
\Delta t_{\rm thermal} \simeq \frac{u_{\rm rad} A_\perp}{\Pi Q^-} = \frac{\delta_{\rm eff}}{2c} \frac{1+\xi_{\rm rad}\tau_{\rm eff}}{1-e^{-\tau_{\rm eff}}} \, .
\end{equation}
For a stationary flow with a large optical depth, this may be further simplified to 
\begin{equation}
\label{E:dt:th2}
  \Delta  t_{\rm thermal} \simeq \frac{\xi_{\rm rad}}{2c} \frac{\varkappa \dot{M}\, \delta_{\rm eff}^2}{A_\perp v}  .
\end{equation}
If the radiation from the lateral sides is ignored, this expression reduces (up to a factor of 4) to equation~(\ref{E:tthermal}). 

The upper limit for the diffusion equation time step is  estimated as (see, for example, the first appendix of ~\citealt{mathphys}):
\begin{equation}\label{E:dt:diff}
    \Delta t_{\rm diff} \simeq C_{\rm diff} \left( \frac{\Delta l^2}{D}\right)_{\rm min},
\end{equation}
where $C_{\rm diff} \leq 0.5$ is a dimensionless factor. 
Overall stability of the numerical solution requires a time step smaller than the smallest of all the three, hence we used the smallest of the three lower limits
\begin{equation}
    \Delta t = \min \left( \min \Delta t_{\rm CFL}, \ \min \Delta t_{\rm thermal},\ \min \Delta t_{\rm diff}\right), 
\end{equation}
where the three time steps are given by equations~(\ref{E:dt:CFL}), (\ref{E:dt:tthermal}), and (\ref{E:dt:diff}).

\section{Results}\label{sec:res}

\subsection{General picture and shock formation}\label{ssec:general_picture}

In Table~\ref{tab:mod}, we summarize the parameters of the models. 
The models have different mass accretion rates, magnetic moments, and azimuthal extents of the flow. 
The mass of the NS is everywhere set to 1.4\Msun, and its radius to $4.86\,GM_*/c^2 \simeq 10\km$. 
For most of the models, we adopt the relative width of the flow at the outer boundary $\Delta R_{\rm e}/R_{\rm e} = 0.25$.
The effective size of the magnetosphere was calculated as $R_{\rm e} = \xim R_{\rm A}$, where $\xim$ is set to 0.5, and $R_{\rm A}$ according is Alfven radius given by equation~(\ref{E:geo:Alfven}). 
For most of the simulations, the size of the magnetosphere is about 14$R_\NS$. 
Coordinate mesh is approximately logarithmic in $l$, with the total number of 9600 radial points for the fiducial model. 
This gives a resolution of about $\Delta l \sim 3\times 10^{-4}R_*$ near the surface of the NS, and $\Delta l \sim 4\times 10^{-3}R_*$ close to the outer edge. 
Most simulations are run for tens of replenishment times, ensuring that mass loss starts well before the end of the simulation. Output is made each 0.01$t_{\rm r}$. 
For model {\tt H}, the output is intentionally made ten times more often to track the dynamical-time-scale variability. 

\begin{table*} 
\caption{ Parameters of the simulations. First column gives the identifier of the model used throughout the paper. 
Subsequent columns are: mass accretion rate normalised by the Eddington value, magnetic moment, azimuthal fraction $a$ occupied by the flow, radius of the magnetosphere, normalised cross-section and latitudinal thickness of the flow near the surface of the star, replenishment time (equation~\ref{E:trep}), duration of the simulation, Basko-Sunyaev's advection parameter $\beta_{\rm BS}$, and essential comments about the particular simulation. 
All the models with $a=1$ have no cooling from the lateral sides.\label{tab:mod}
}
\small
\begin{tabular}{lcccccccccp{3cm}}
\hline \hline
ID & $\displaystyle \frac{\dot{M}_{\rm out}c^2}{L_{\rm Edd}}$ & $\mu_{30}$,  & $a$ & 
$R_{\rm e}/R_*$ & $A_\perp(R_*)/R_*^2$ & $\delta(R_*)/R_*$ & $t_{\rm r}$, s & $t_{\rm max}$, s & $\beta_{\rm BS}$  & comment\\
   & & $10^{30}\gcmc$ \\
\hline
F   & $10$   & 0.1 & 0.25 & 13.95 & $2.89\times 10^{-2} $ & 0.0344 & 0.11 & 1  & 0.43 & fiducial \\ 
L & $10$   & 0.1 & 0.25 & 13.95 & $2.89\times 10^{-2} $ & 0.0344 & 0.11 & 1.4 & 0.43 & 2X coarser resolution\\ 
F2   & $10$   & 0.1 & 0.25 & 13.95 & $2.89\times 10^{-2} $ & 0.0344 & 0.11 & 1  & 0.43 & 2X finer resolution \\ 
ND   & $10$   & 0.1 & 0.25 & 13.95 & $2.89\times 10^{-2} $ & 0.0344 & 0.11 & 2  & 0.43 & no radial photon diffusion \\ 
B   & $10$   & 0.1 & 0.25 & 13.95 & $2.89\times 10^{-2} $ & 0.0344 & 0.11 & 2  & 0.43 &   no radial photon diffusion; no radiation losses from lateral sides \\ 
M1   & $1$   & 0.03 & 0.25 & 13.53 & $2.985\times 10^{-2} $ & 0.0350 & 0.10 & 1.1  & 0.10 & \\ 
M3   & $3$   & 0.05 & 0.25 & 13.24 & $3.05\times 10^{-2} $ & 0.0353 & 0.10 & 0.6  & 0.22 & \\ 
M30  & $30$ & 0.2 & 0.25 & 14.19 & $2.66\times 10^{-2}$ & 0.0330 & 0.14 & 2  &  0.64 &  \\
M100  & $100$ & 0.3 & 0.25 & 13.53 & $2.99\times 10^{-2}$ & 0.0345 & 0.10 & 2  &  0.81 &  \\
W   & $10$   & 0.1 & 1 & 13.95 & $0.116$ & 0.0344 & 0.43 & 2 & 0.45  & no cooling from the lateral sides \\ 
N  & $10$ & 0.1 & 0.05 & 13.95 & $5.79\times 10^{-3}$ & 0.0344 & 0.022 & 0.38  &  0.73 &  \\
N2  & $10$ & 0.1 & 0.05 & 13.95 & $5.79\times 10^{-3}$ & 0.0344 & 0.022 & 0.38  &  0.73 & no cooling from the lateral sides\\
R   & $10$   & 0.1 & 0.25 & 13.95 & $2.89\times 10^{-2} $ & 0.0344 & 0.11 & 1.9 & 0.43 &  $\Omega = 0.9\Omega_{\rm K}(R_{\rm e})$\\ 
I   & $10$   & 0.1 & 0.25 & 13.95 & $2.89\times 10^{-2} $ & 0.0344 & 0.11 & 0.9 & 0.43 &  irradiation $\eta_{\rm irr}=0.5$\\ 
WI   & $10$   & 0.1 & 1 & 13.95 & $0.116$ & 0.0344 & 0.43 & 1.8 & 0.45  &  same as W but $\eta_{\rm irr} = 0.5$\\ 
WI1   & $10$   & 0.1 & 1 & 13.95 & $0.116$ & 0.0344 & 0.43 & 1.9 & 0.45  &  same as W but $\eta_{\rm irr} = 1$\\ 
RI   & $10$   & 0.1 & 0.25 & 13.95 & $2.89\times 10^{-2} $ & 0.0344 & 0.11 & 0.8 & 0.43 &  $\Omega = 0.9\Omega_{\rm K}(R_{\rm e})$, $\eta=0.5$\\ 
H & $10$ & 1 & 0.25 & 52.0 & $7.61\times 10^{-3} $ & 0.0175 & 2.9 & 1.4  &  0.417&  \\
M100W2x & $100$ & 0.3 & 1.0  & $23.7$ & $0.118$ & 0.0488 & 0.4  & 3.5 & 0.72 & $\Delta R_{\rm e}/R_{\rm e} = 0.5$, $\xim = 1$, 2X coarser resolution\\
M100W3  & $100$ & 0.3 & 1.0 & 13.53 & $0.143$ & 0.0419 & 0.50  & 2  & 0.63 &  $\Delta R_{\rm e}/R_{\rm e} = 0.3$\\
M100W4  & $100$ & 0.3 & 1.0 & 13.53 & $0.119$ & 0.0350 & 0.42 & 2  & 0.60 &  \\
M100W5  & $100$ & 0.3 & 1.0 & 13.53 & $0.0955$ & 0.0280 & 0.33 & 2  & 0.56 &  $\Delta R_{\rm e}/R_{\rm e} = 0.2$\\
M100W10  & $100$ & 0.3 & 1.0 & 13.53 & $0.0478$ & 0.0140 & 0.17 & 3  & 0.43 &  $\Delta R_{\rm e}/R_{\rm e} = 0.1$\\
M100W20  & $100$ & 0.3 & 1.0 & 13.53 & $0.0239$ & 0.00700 & 0.08 & 1.4  & 0.308 &  $\Delta R_{\rm e}/R_{\rm e} = 0.05$\\
M100W50  & $100$ & 0.3 & 1.0 & 13.53 & $0.00955$ & 0.00280 & 0.033 & 0.6  & 0.178 &  $\Delta R_{\rm e}/R_{\rm e} = 0.02$\\
\end{tabular}
\end{table*}

Table~\ref{tab:mod} also contains the quantity $\beta_{\rm BS}$, that is the combination of global parameters (geometry, magnetic moment, and mass accretion rate) that determines the fraction of the accretion luminosity advected towards the NS surface in the solution of \citetalias{BS76}. 
To calculate this parameter, we used the expressions of section 4.4 of \citetalias{BS76}. 
Detailed comparison with the analytic solution will be done later in section~\ref{sec:res:compare}.

From the beginning, the flow, initially practically  static (velocities much smaller than Keplerian by absolute values), starts falling along the field lines. 
The infall leads to adiabatic heating and subsequently to the formation of a shock near the bottom of the simulation domain.
The shock moves upward, approaching the equilibrium position even before the beginning of the mass loss.
In the steady state, the flow in the column is heated by adiabatic compression, while the excess thermal energy is radiated away. 
Energy release and radiation loss evolve towards the equilibrium on the time scale close to replenishment time (equation~\ref{E:trep}) or the thermal time at the bottom of the column (equation~\ref{E:tthermal:bottom}). 

\begin{figure*}
\includegraphics[width=0.9\textwidth]{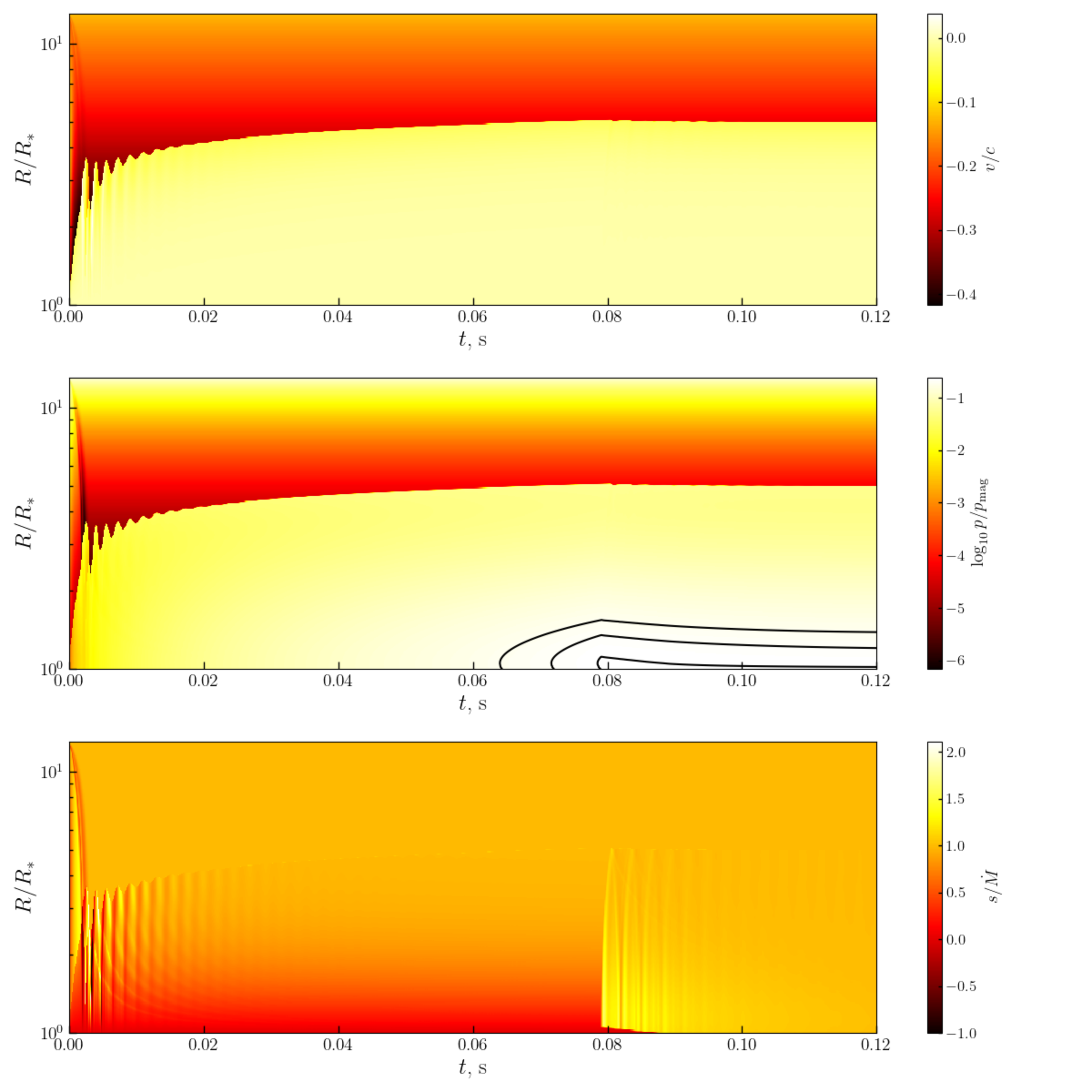}
 \caption{Time-radius diagrams for velocity (upper panel), thermal pressure (middle panel; the logarithm of the quantity in $\pmag$ units is shown), and mass flux $s$ (lower panel; the quantity is normalised by the mass accretion rate at the outer boundary) for the first 0.12\,s of simulation {\tt N}. 
  Black contours in the middle panel correspond to pressure equal to 80, 90, and 99 per cent of the magnetic field pressure. 
 }\label{fig:start_narrow}       
\end{figure*}

In all the simulations, the shock oscillates before reaching the steady state.
The oscillations are damped, though it is difficult to decide if the damping is physical or numerical. 
We discuss these oscillations in more detail in section~\ref{sec:res:var}. 
In Fig.~\ref{fig:start_narrow}, we show the development of the shock wave on the time-radius plots for velocity and thermal energy density for model {\tt N}. 
In the velocity plot, the shock wave is clearly seen as the boundary between the almost static region closer to the surface of the star (column) and the nearly free-falling region above.

In Table~\ref{tab:res}, we give the position of the shock and different characteristic luminosities emitted by the flow after it reaches a steady state. 
The luminosity of the flow is integrated using equation~(\ref{E:ltot}) over the whole simulation domain ($L_{\rm tot}$) and below the shock wave ($L_{\rm X}$). 
$L_{\rm out}$ is the thermal power entering from the outer border (see equation~\ref{E:lout}).
The last column in Table~\ref{tab:res} gives the fraction of the luminosity that is not radiated by the accretion column but rather lost with the mass.
In our setup, its role is the same as of the advected power fraction in \citetalias{BS76}. 
For most of the models, $1- L_{\rm X}/L_{\rm acc} \simeq \beta_{\rm BS}$, and the deviations are mostly related to the emission from the lateral sides or to the heat flux at the outer boundary (as in {\tt M3}  where  the total luminosity is larger than $L_{\rm acc}$). 

\begin{table*}\centering
\caption{ Parameters of the column measured for the last $10$ per cent of the simulation time for different models. 
Luminosities $L_{\rm tot}$ and $L_{\rm X}$ are calculated by integrating the radiation flux over the whole simulation and below the shock, respectively. 
`Predicted' shock position and luminosity are estimated in the framework of the analytic model by \citetalias{BS76}.
Luminosities $L_{\rm acc}$, $L_{\rm tot}$, $L_{\rm X}$, and $L_{\rm out}$ are defined in sections~\ref{sec:num:con} and \ref{sec:num:BC}.
}
\label{tab:res}
\begin{tabular}{lccccccccccc}
\hline \hline
ID & \multicolumn{2}{c}{$R_{\rm shock} / R_*$}  & ~~~ &  $L_{\rm tot}/L_{\rm Edd}$ & $L_{\rm acc}/L_{\rm Edd}$ &~~~ & \multicolumn{2}{c}{$L_{\rm X}/L_{\rm Edd}$} & $L_{\rm out}/L_{\rm Edd}$ & $1-\frac{L_{\rm X}}{L_{\rm acc}}$\\
& measured & predicted  &&    &  &&measured &predicted\\
\hline
F & 3.238$\pm$0.005  & 3.58 && 1.44  & 2.06 && 1.16 & 1.17 & 0.38 & 0.44\\
L & 3.238$\pm$0.009  & 3.58 && 1.44  & 2.06 && 1.16 & 1.17 & 0.38 &0.44\\
F2 & 3.237$\pm$0.002  & 3.58 && 1.42  & 2.06 && 1.17 & 1.17 & 0.38 &0.43\\
ND & 3.260$\pm$0.005  & 3.58 && 1.33  & 2.06 && 1.18 & 1.17 & 0.38 & 0.43 \\
B & 3.567$\pm$0.005  & 3.58 && 1.27  & 2.06 && 1.12 & 1.17 &0.38 & 0.46\\
M1 & 1.3724$\pm$0.0019 & 1.42 && 0.201  & 0.205 && 0.180 & 0.184  & 0.21 & 0.12\\
M3 & 1.837$\pm$0.004 & 1.99 && 0.70  & 0.53 && 0.66& 0.48 & 0.21 & -0.24\\
M30 & 6.506$\pm$0.010 & 7.39 && 2.70  & 6.15 && 2.16& 1.16 & 0.72 & 0.65\\
M100 & -- & 17.87 && 3.67  & 20.5 && --& 3.89 & 2.3 & --\\
W & 1.980$\pm$0.003  & 1.97 && 2.87  & 2.06 && 1.63 & 1.65 & 1.30 & 0.21\\
N & 4.982$\pm$0.007 & 8.47 && 0.81  & 2.06 && 0.72& 0.55 & 0.13 & 0.65\\
N2 & 8.447$\pm$0.04 & 8.47 && 0.57  & 2.06 && 0.51& 0.55 & 0.13 & 0.75\\
R & 3.137$\pm$0.004  & 3.58 && 1.25  & 2.06 && 1.09 & 1.17 & 0.38 & 0.47\\
I & 3.256$\pm$0.005  & 3.58 && 1.34  & 2.06 && 1.19 & 1.17 & 0.38 & 0.42\\
WI & 1.962$\pm$0.003  & 1.97 && 2.8  & 2.06 && 1.55 & 1.65 & 1.29 & 0.25\\
WI1 & 1.98$\pm$0.04  & 1.97 && 2.05  & 2.06 && 1.64 & 1.65 & 1.33 & 0.20 \\
RI & 3.215$\pm$0.005  & 3.58 && 1.31  & 2.06 && 1.16 & 1.17 & 0.38 & 0.44\\
H & 3.786$\pm$0.004 & 4.40 &&1.31 & 2.05 &&  1.26 & 1.20 & 0.47 & 0.39\\
M100W2x & 10.73$\pm$0.04 & 12.1 && 6.35 & 20.5 && 5.59 &  5.74 & 2.19 & 0.73\\
M100W3 & 12.450$\pm$0.010 & 8.44 && 10.8 & 20.5 && 9.9 & 7.6  & 6.5 & 0.52\\
M100W4 & 8.869$\pm$0.016 & 7.27 && 12.8 & 20.5 && 9.2 & 8.3  & 6.1 & 0.55\\
M100W5 & 6.485$\pm$0.009 & 6.09 && 12.8 & 20.5 &&  9.1 & 9.2  & 5.3 & 0.56 \\
M100W10 & 3.733$\pm$0.005 & 3.71 && 13.7 & 20.5 &&  11.5 & 11.7  & 3.15 & 0.44\\
M100W20 & 2.458$\pm$0.003 & 2.47 && 15.2  & 20.5 && 13.8 &  14.2  & 1.97 & 0.33\\
M100W50 & 1.656$\pm$0.002 & 1.66 && 17.3  & 20.5 && 16.1 & 16.7  & 1.23 & 0.21\\
\end{tabular}
\end{table*}

\begin{figure}
\adjincludegraphics[width=\columnwidth,trim={0cm 0.4cm 0cm 0cm},clip]{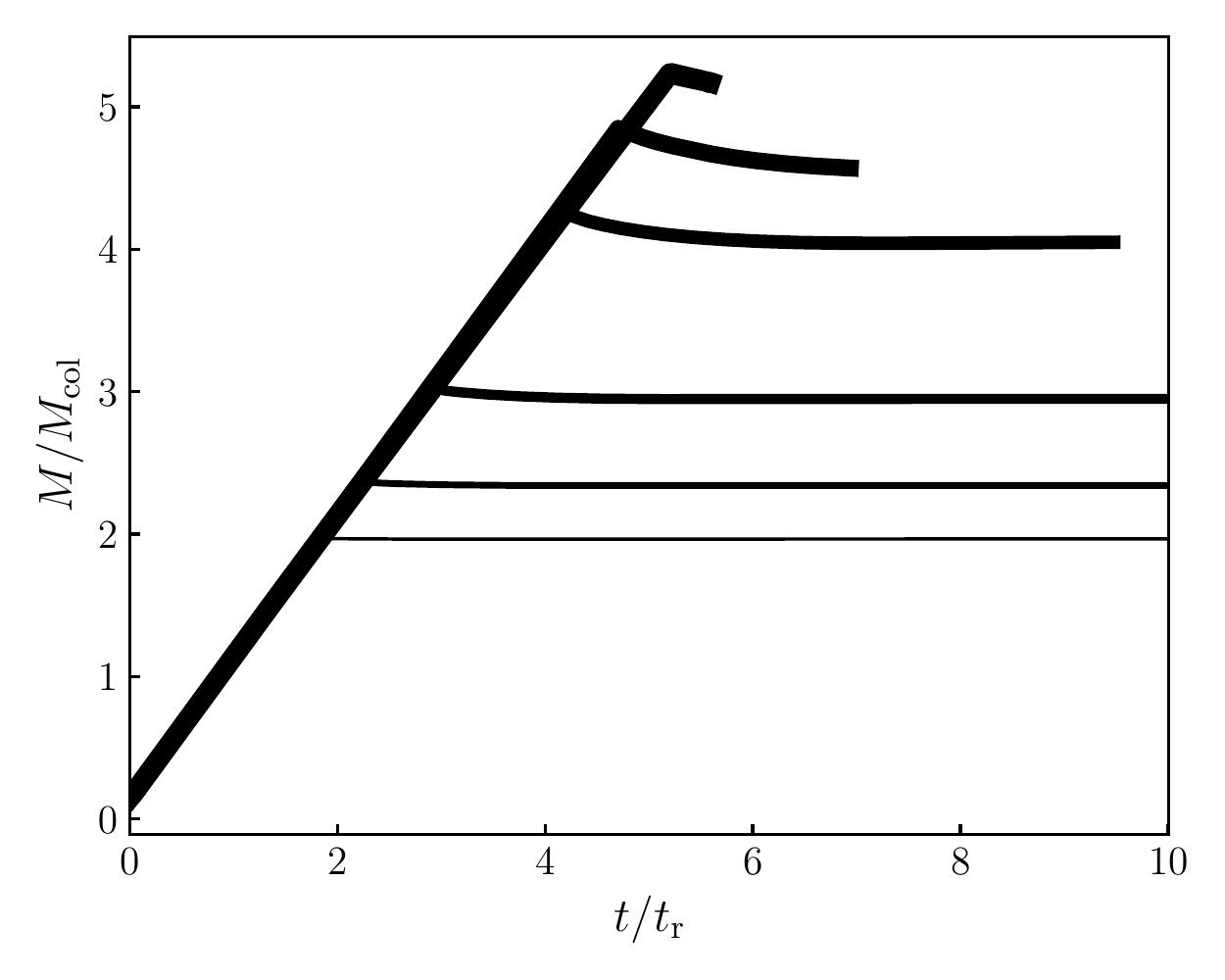}
 \caption{ Total mass (in the estimated column mass units, see equation~\ref{E:mcol}) as a function of time (replenishment time units, equation~\ref{E:trep}) for models {\tt M100W3-50}, from the thickest to the thinnest curve. 
 }\label{fig:massrace}       
\end{figure}

\subsection{Mass leakage from the column}

After the position of the shock stabilises, mass starts being lost from the column.
The higher the position of the shock front, the higher the place where thermal pressure starts exceeding magnetic pressure.
In general, the region where the mass leaks from the column (that we will hereafter refer to as the \emph{vent}) is formed above the surface of the star, sometimes at a height comparable to the radius of the NS. 

In Table~\ref{tab:vents}, we give the radial coordinates at which the vents first appear in the models where the vent first appears at a considerable altitude.
The initial vent position depends not only on the accretion rate but also on the geometry of the flow.
In particular, for model {\tt N2} (Fig.~\ref{fig:start_narrow}), the critical energy density is reached at about 2$R_*$, though the mass accretion rate is the same as in model {\tt F}, where the vent appears very close to the surface. 
In most of the cases (with the exception of  models {\tt N2}, {\tt M100} and {\tt M100W2x}, considered in more detail later in section~\ref{sec:res:breakdown}) when the vent is formed well above the surface, it gradually drifts downwards and eventually reaches the surface of the star. 

There are several parameters that define the geometry of an accretion column.
Small radiating perimeter $\Pi$ (as in models {\tt N} and {\tt N2}) decreases the radiating surface, thereby trapping more heat inside the column. 
This leads to a higher shock and a higher initial vent, especially in {\tt N2} where the radiation losses from the lateral sides of the column are ignored. 
Decreasing the latitudinal size of the column (by decreasing $\Delta R_{\rm e}$), on the other hand, facilitates cooling of the column by decreasing the average optical depth, and the mass leakage starts closer to the surface. 

Very tall columns like that in {\tt N2} may have several vents at different heights in the steady-state regime. 
For {\tt M100}, where no stationary shock wave develops, mass loss occurs over a broad range of heights, approximately from $1$ to $2.5-3\,R_\NS$. 
We will discuss the behaviour of these models in more detail in section~\ref{sec:res:breakdown}.
To sum up, same factors (accretion rate and flow geometry) affect the height of the column (the shock position) and the initial vent position. 

\begin{figure*}
\includegraphics[width=0.9\textwidth]{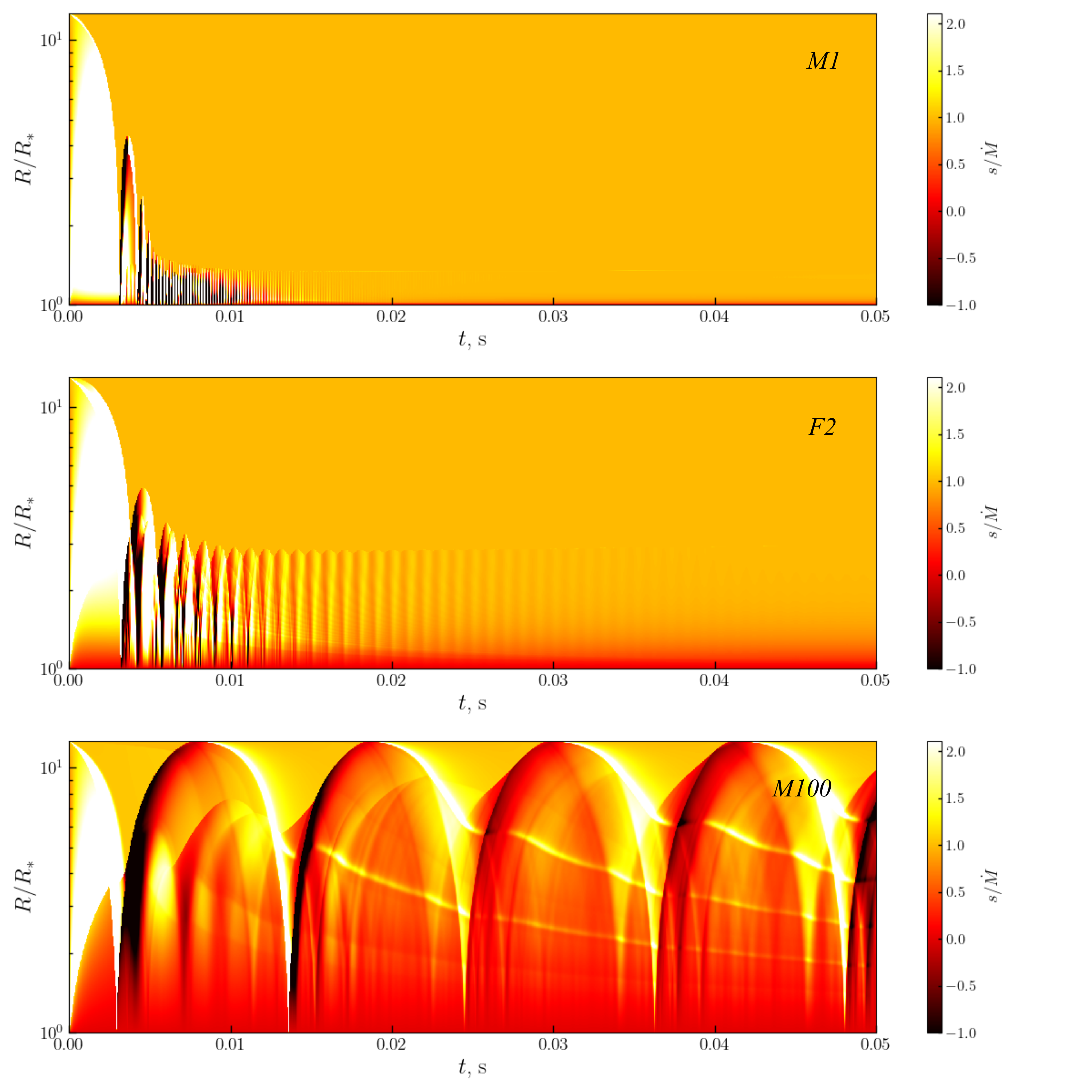}
 \caption{ Local mass accretion rate plotted as a function of radius and time for three models ({\tt M1}, {\tt F}, and {\tt M100}). Only the first 0.05s of evolution are shown. }\label{fig:mdotmaps}
\end{figure*}

In Fig.~\ref{fig:mdotmaps}, we show the evolution of the mass flux during the initial 0.05\,s (about half of the replenishment time) for three different models.
While above the shock, the mass accretion rate $s = \rho v A_\perp$ is very stable and close to $\dot{M}$, matter inside the column is, for most of the models, involved in a cyclic motion with the velocity amplitude exceeding the mean velocity value. 
Initial shock oscillations leave entropy traces in the flow, that are best visible in the diagram for {\tt M100}. 
In {\tt N2} and in {\tt M100}, the oscillations do not show any damping, and the position of the vent neither stabilises nor approaches the surface of the star. 
In {\tt M100W2x}, the oscillations are damped significantly, and the flow structure stabilises, with the position of the vent at a height of about 0.2$R_\NS$ above the NS surface. 
The damping may be a consequence of the lower resolution of this model.

\begin{figure*}
\adjincludegraphics[width=0.9\textwidth,trim={0cm 0.4cm 0cm 0cm},clip]{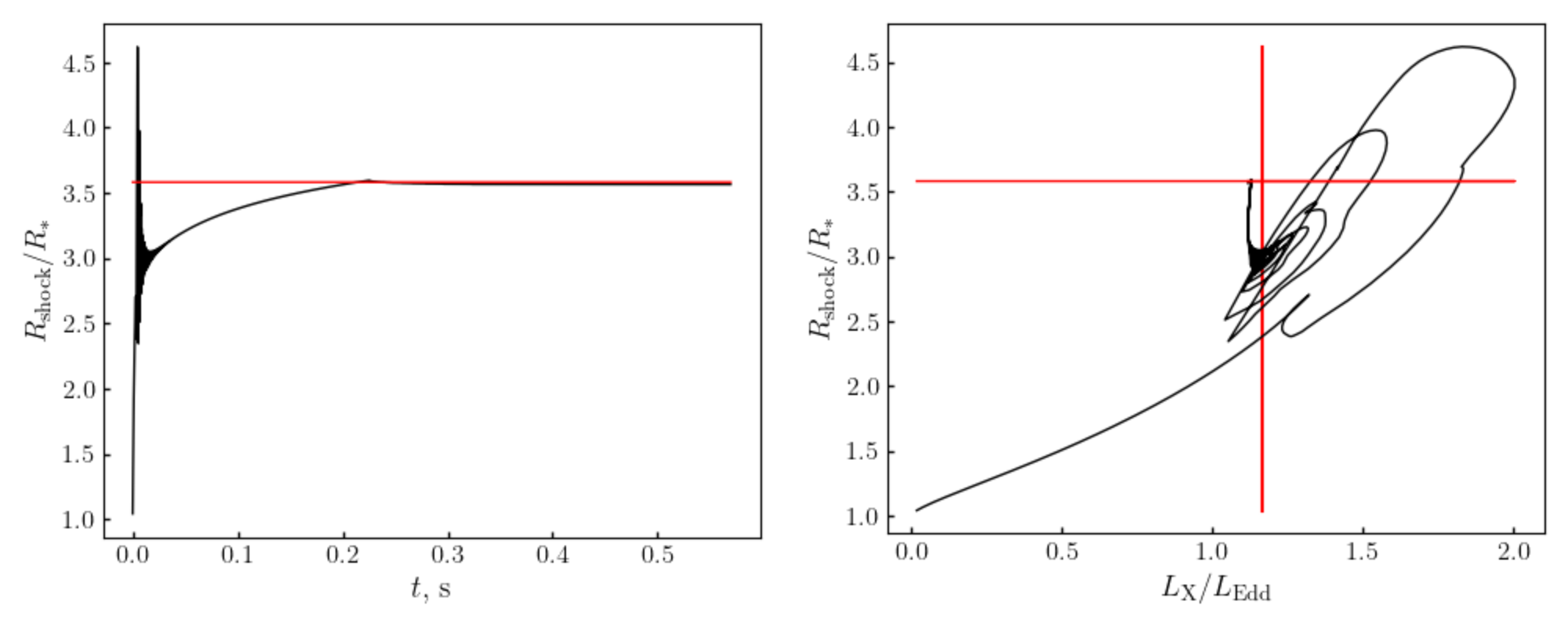}
 \caption{Shock position as a function of time (left panel) and luminosity $L_{\rm X}$ emitted below the shock (right panel) for model {\tt B}. 
 Shock position predicted by \citetalias{BS76} is shown with a horizontal red line, and the predicted luminosity is shown with a vertical red line in the right panel. }\label{fig:sfronts}       
\end{figure*}

\begin{table}\centering
\caption{ 
The initial position of the vent  and the time of its opening for different models.
}
\label{tab:vents}\small 
\begin{tabular}{lcc}
\hline \hline
ID & $R_{\rm vent}/R_*$ & $t_0$, ms\\
\hline
F & $1.00\pm0.0003$ & $207.52\pm 0.04$ \\
F2 & $1.00\pm0.0003$ & $207.86\pm 0.04$ \\
B & $1.0\pm 0.0003$ & $233.80\pm 0.04$ \\
N & $1.0538\pm 0.0003$ & $79.10\pm 0.08$\\
N2 & $1.9096\pm 0.0005$ & $96.18\pm 0.08$\\
M30 & $1.0381\pm 0.0003$ & $466.0\pm 0.5$ \\
M100 & $11.263\pm 0.002$ & $6.20\pm 0.04$ \\
M100W2x & $1.5430\pm 0.0010$ & $813.3\pm 0.4$\\
M100W3 & $1.0808\pm 0.0002$ & $1909\pm 2$ \\
M100W4 & $1.0229\pm 0.0003$ & $1380.0\pm 1.4$ \\
M100W5 & $1.0122\pm 0.0003$ & $946\pm 11$ \\
M100W10 & $1.0003\pm 0.0003$ & $344.1\pm 0.6$ \\
M100W20 & $1.0\pm 0.0003$ & $141.1\pm 0.3$ \\
M100W50 & $1.0\pm 0.0003$ & $81.78\pm 0.11$ \\
\end{tabular}
\end{table}

\subsection{Structure of the flow and comparison with the analytical model}\label{sec:res:compare}

The shock front position is easy to find as the position of the velocity derivative maximum.
In Fig.~\ref{fig:sfronts}, we show the  position of the shock front as a function of time and below-the-shock luminosity $L_{\rm X}$ for the fiducial model {\tt F}.
In simulation {\tt F}, mass starts to leak out at $t\sim 0.2$\,s, which  is visible in the left panel of Fig.~\ref{fig:sfronts} as a break in the shock front motion law.

We compare the position of the shock with the prediction of equation 34 of \citetalias{BS76}
\begin{equation}\label{E:BS:xs}
    \displaystyle \eta \gamma^{1/4} \xi_{\rm s}^{n/4+1/8} = 1+{\rm e}^{\gamma \xi_{\rm s}} \left[ \xi_{\rm s} \eint{2}{\gamma} - \eint{2}{\gamma \xi_{\rm s}}\right],
\end{equation}
that implicitly defines the radial coordinate of the shock $\xi_{\rm s} = R_{\rm shock}/R_{*}$. 
Here, $n$ is a dimensionless parameter describing the curvature of the field lines (for dipolar geometry, $n=3$), and 
\begin{equation}
    \eint{k}{x} = \int_1^{+\infty} t^{-k} {\rm e}^{- t x} \diff t
\end{equation}
is the exponential integral of order $k$. 
Dimensionless coefficients $\gamma$ and $\eta$ in equation~(\ref{E:BS:xs}) are
\begin{equation}
\label{E:BS:gamma}
\begin{array}{l}
   \displaystyle \gamma = \frac{c\,  R_*}{\varkappa  \delta^2(R_*) } \frac{A_{\perp}(R_*)}{\dot{M}}\, \frac{3}{2\xi_{\rm rad}} \\
  \displaystyle   \qquad{} \simeq 0.387 \ \frac{A_{\perp}(R_\NS)}{ \delta^2(R_\NS)}\, \frac{L_{\rm Edd}}{\dot{M}_{\rm out} c^2}\, \frac{R_\NS c^2}{4.86 GM_\NS}\, \frac{3}{2\xi_{\rm rad}} \\
    \end{array}
\end{equation}
and 
\begin{equation}
\label{E:BS:eta}
\begin{array}{l}
    \displaystyle \eta = \left(\frac{8\varkappa}{21c} \frac{3\pmag(R_\NS)\,\delta^2(R_\NS)}{\sqrt{2GM_\NS R_\NS}} \frac{2\xi_{\rm rad}}{3}\right)^{1/4} \\
    \displaystyle \simeq 12.57 \ \left( \frac{B}{10^{12}\rm G}\right)^{1/2} \left( \frac{\delta(R_\NS)}{0.03 R_\NS}\right)^{1/2}
    \left( \frac{R_\NS c^2}{4.86GM_\NS} \right)^{3/8} \left( \frac{M}{1.4\Msun} \right)^{1/4}.\\
    \end{array}
\end{equation}
In comparison with the original expressions of \citetalias{BS76}, equations~\eqref{E:BS:gamma} and \eqref{E:BS:eta} contain an extra factor of $2\xi_{\rm rad} /3$, resulting from the radiation diffusion across the flow, parameterised by the coefficient $\xi_{\rm rad}$. 
Here, we set $\xi_{\rm rad} =3/2$, that corresponds to all the energy released in the middle of the column.
The position of the shock may be approximated as $\xi_{\rm s} \simeq 1 + \ln \left[ \eta \gamma^{1/4} (1+\gamma)\right]\big/\gamma$ \citep{Lyubarskii-Syunyaev1988} to an accuracy about $10\%$ for $\gamma\gtrsim 0.1$.

Position of the shock $\xi_{\rm s}$ and the two parameters $\gamma$ and $\eta$ may be used to find the advection parameter as 
\begin{equation}
\label{E:BS:beta}
    \beta_{\rm BS} = 1 - \gamma {\rm e}^\gamma \left[ \eint{1}{\gamma} - \eint{1}{\gamma \xi_{\rm s}}\right].
\end{equation} 
The luminosity of the column itself is equal to $\left( 1-\beta_{\rm BS}\right)L_{\rm acc}$.
With few exceptions, we find this estimate to work well for our simulations after they enter the steady-state regime (see section~\ref{ssec:general_picture} and Tables~\ref{tab:mod} and \ref{tab:res}). 

\begin{figure}
\adjincludegraphics[width=0.9\columnwidth,trim={0cm 0.4cm 0cm 0cm},clip]{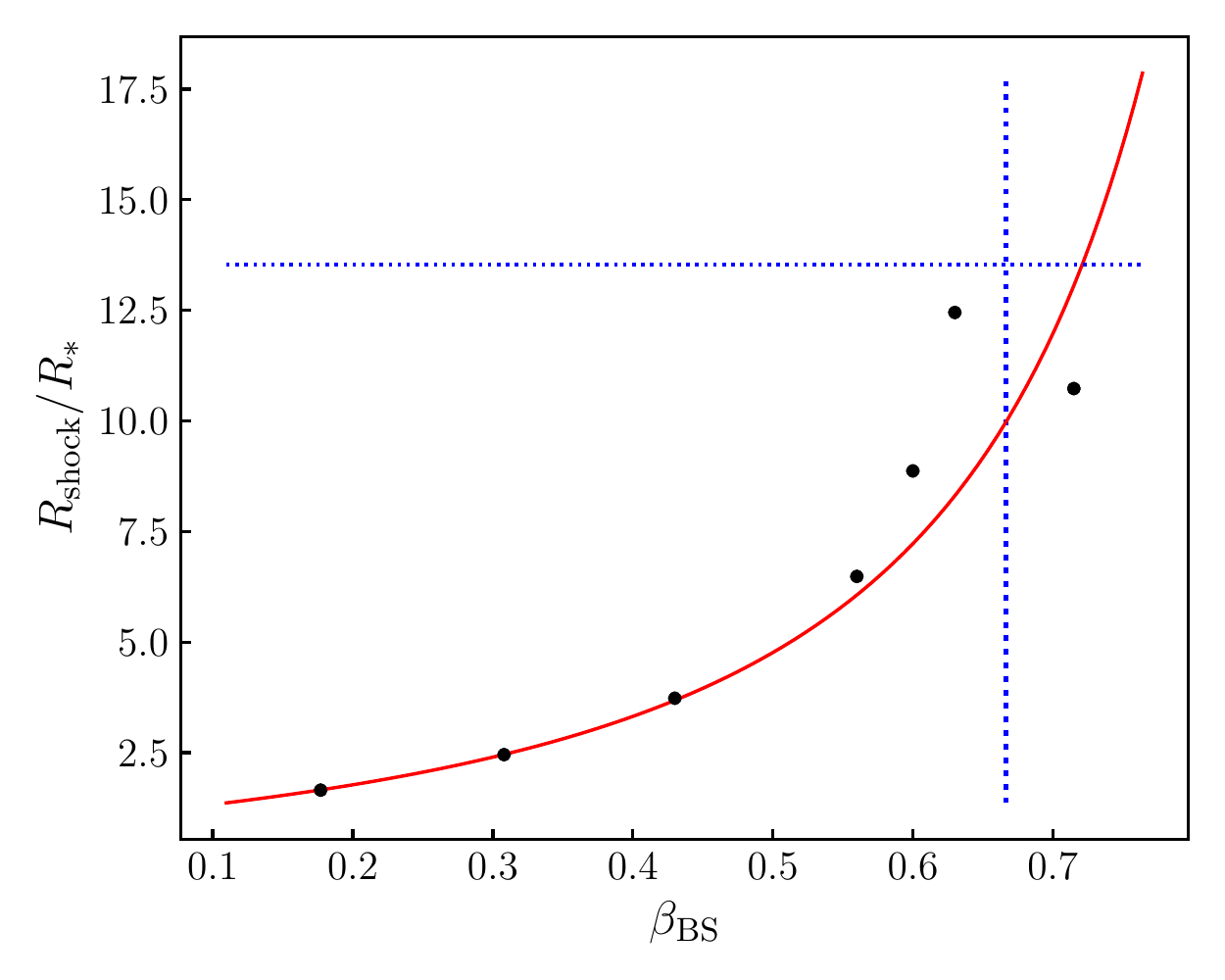}
 \caption{Shock front position as a function of $\beta_{\rm BS}$, shown for models {\tt M100W2x} and {\tt M100W3-50}, differing in $\Delta R_{\rm e}/ R_{\rm e}$ (black dots).
Predictions of the analytic model (equations~\ref{E:BS:xs} and \ref{E:BS:beta}) are shown with a red line.
 Vertical and horizontal blue dotted lines show the critical value $\beta_{\rm BS} = 2/3$ and the size of the magnetosphere. 
 }\label{fig:betax}
\end{figure}

In Table~\ref{tab:res}, we give the shock positions predicted by \citetalias{BS76} as well as the simulation results. 
For {\tt B} and several other models, the prediction is pretty accurate (about 0.4 per cent for {\tt B}). 
Most of the difference between predicted and calculated quantities in Table~\ref{tab:res} comes from the radiation losses from the lateral sides of the column. 
Also, the higher the shock position, the less accurate is the analytic approximation, as the latter does not account for the real dipole geometry.  
The shock position is not very sensitive to the effects of photon diffusion (compare {\tt F} and {\tt ND}).
Other physical effects also affect only slightly the shock position, as we demonstrate later in section~\ref{sec:res:eff}.

Our fiducial model uses physical assumptions different from that of the analytic model by \citetalias{BS76}. 
Apart from time dependence, we use dipole geometry, different set of boundary conditions, and allow for photon diffusion along magnetic field lines. 
In model {\tt B}, photon diffusion is turned off, and the contribution of the lateral sides of the column to radiation losses is ignored. 
This conforms well to the assumptions of \citetalias{BS76}, and allows straightforward comparison to the analytic solution. 
Quite expectedly, the shock position in {\tt B} is very close to the predicted value.

 The effect of changing geometry, as well as consistency with \citetalias{BS76}, is illustrated by Fig.~\ref{fig:betax}, where we show six models {\tt M100W3-50}, differing only in the width of the penetration zone, and an additional model {\tt M100W2x}, having similar geometry but a larger magnetosphere.
For a fixed mass accretion rate, smaller $\delta$ means higher surface-to-volume ratio, more efficient cooling and thus lower shock and vent positions. 
For small advection parameters $\beta_{\rm BS} \lesssim 0.6$, there is good agreement between our simulations the analytic results, while the last three points deviate from the analytic model considerably.

\begin{figure}
\adjincludegraphics[width=1.0\columnwidth,trim={0cm 0.25cm 0cm 0cm},clip]{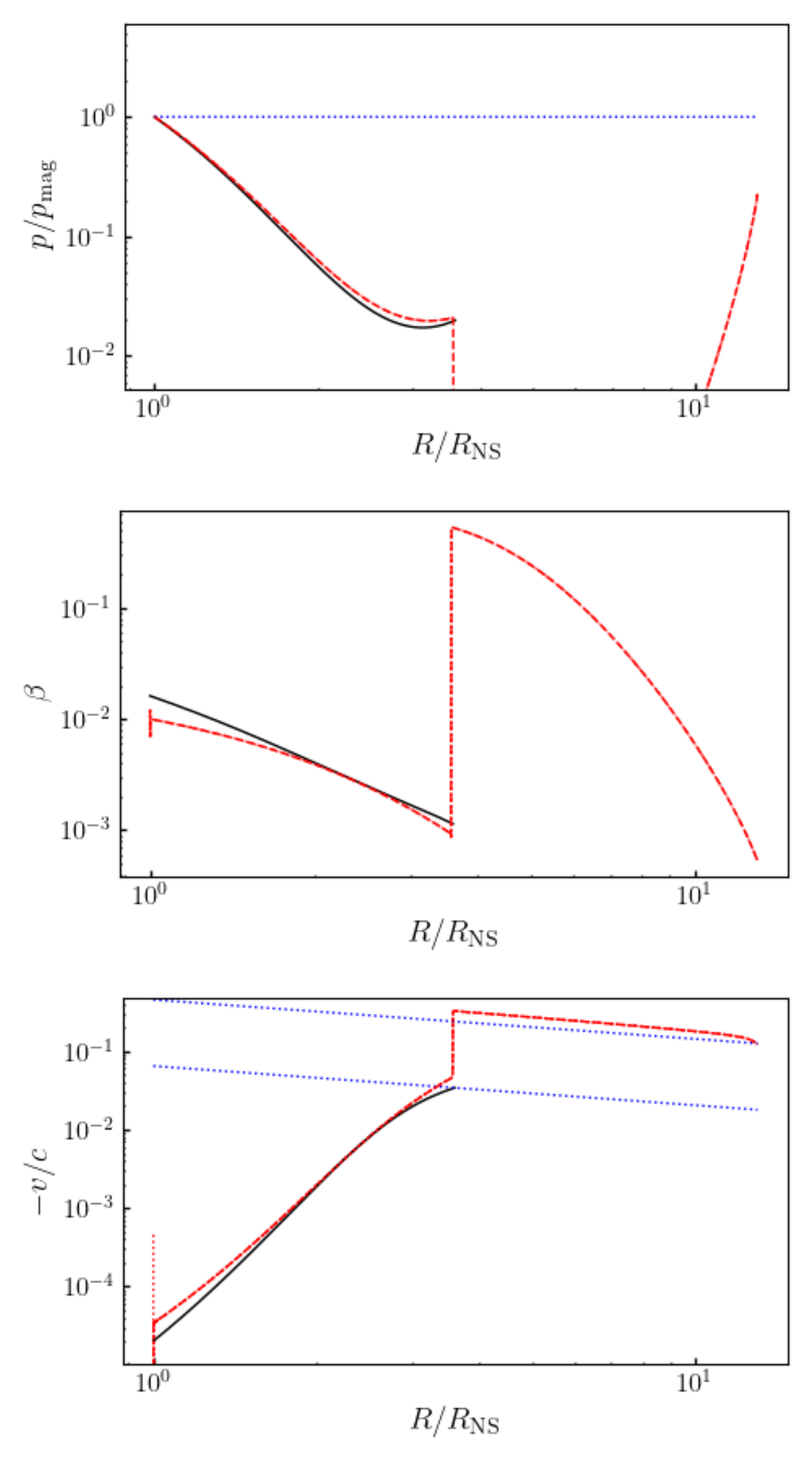}
 \caption{
 Stationary analytic solution (black solid lines) compared with our model {\tt B}, averaged over the period of time $1.1-1.9$\,s.
  Upper panel: pressure normalised by the local magnetic field pressure. Blue dotted line shows the $p =  \pmag$ condition. Middle panel: gas-to-total pressure ratio. Lower panel: radial velocity. Blue dotted lines in the lower panel are Keplerian velocity $v_{\rm K}$ and $v_{\rm K}/7$. 
 }\label{fig:BScompare}
\end{figure}

While the position of the shock  is generally well predicted by the analytical solution, the differences in the detailed column structure are more pronounced (see Figs~\ref{fig:BScompare} and \ref{fig:BScompare:N}).
In Fig.~\ref{fig:BScompare}, we compare time-averaged properties of the simulated flow with the analytic solution. 
Certain differences arise from the boundary conditions near the surface of the NS. 
While the analytic model requires a non-zero velocity at $R=R_\NS$, our model allows the matter to seep out at a zero radial velocity. 
Existence of such a sink creates a velocity discontinuity near the surface, above which the flow moves at a speed even higher than predicted by the analytic model. 

Another difference between our results and analytic predictions is in the contribution of the gas pressure. 
While $p_{\rm gas}$ is still much smaller than $p_{\rm rad}$ everywhere below the shock, the value of $\beta = p_{\rm gas}/p$ differs from the analytic model by up to a factor of 2 near the surface. 

\subsection{Breakdown at large accretion rates}\label{sec:res:breakdown}

\begin{figure}
\adjincludegraphics[width=1.0\columnwidth,trim={0cm 0.25cm 0cm 0cm},clip]{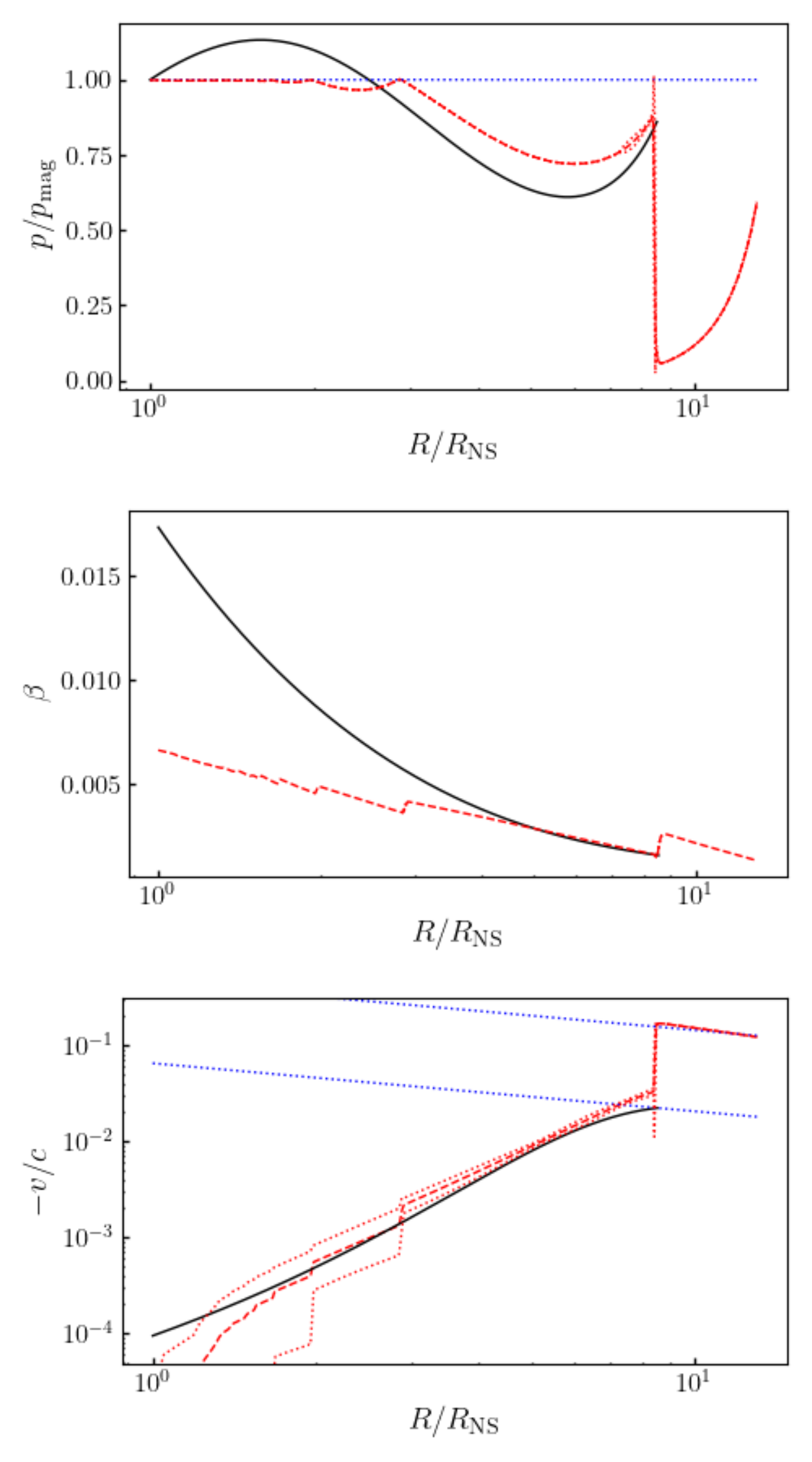}
 \caption{
 The same as previous figure, but for the model {\tt N2}. Averaging was done over the time range $0.3-0.38$\,s. 
 Red dotted lines show the variation limits of the time-dependent solution (mean value plus/minus one standard deviation).
 }\label{fig:BScompare:N}
\end{figure}

For some combinations of parameters, the steady-state analytic solution is physically inconsistent. 
This is generally the case for the parameter values predicting a very tall column, like those of our models {\tt N2} (see Fig.~\ref{fig:BScompare:N}), {\tt M100}, and {\tt M100W2x}. 
For {\tt M100}, the predicted height of the column is larger than the size of the magnetosphere, hence we do not expect any agreement in the structure of the column. 

For {\tt N2}, the analytic solution predicts radiation pressure higher than magnetic pressure at a finite height above the NS surface.
Because of our assumptions about the mass loss from the column,  pressure in the simulation practically never exceeds the pressure of the magnetic field (see Fig.~\ref{fig:BScompare:N}). 
Instead, the column leaks over a broad range of heights, up to about $3\, R_\NS$. 
The position of the shock oscillates with a small but measurable amplitude (about 0.04$R_\NS$, see Table~\ref{tab:res}) around the value predicted by the analytic model. 
The time-averaged structure of the column is also reasonably consistent with the predictions of \citetalias{BS76}.

In models {\tt N2} and {\tt M100}, the position of the shock, as well as the position of the vent, does not approach a steady-state value and oscillates with the period of the pulsations which we discuss in section~\ref{sec:disc:osc}.
In {\tt M100W2x}, the positions of both the shock and the vent stabilise with time, but the vent is located well above the surface.

A sufficient condition for the
analytic solution to become self-inconsistent is a shallow enough radial dependence of $p(R)$ near $R=R_*$. 
If
\begin{equation}
    \left. \frac{\diff \ln p }{ \diff \ln R }\right|_{ R = R_\NS} 
    >\left. \frac{\diff \ln \pmag }{ \diff \ln R }\right|_{ R = R_\NS},
\end{equation}
the boundary condition $p=\pmag$ at the surface of the star implies $p>\pmag$ somewhere above the surface. 
For dipolar magnetospheres, $\diff \ln p_{\rm mag} / \diff \ln R \simeq -6$.
One can check that for the solution given by equation (32) of \citetalias{BS76}, this is equivalent to $\beta_{\rm BS} >2/3$. 

As $\beta_{\rm BS}$ has the physical meaning of the power fraction trapped within the accretion column, large values of $\beta_{\rm BS}$ mean inefficient radiation losses from its surface.
Inefficient cooling of the column not only increases its supply of thermal energy, but also leads to a shallow energy density profile, and, as a consequence, overheating of the column at a finite height.
Indeed, in an advection-dominated column (in the limit $\beta_{\rm BS} \to 1$), the power advected vertically at some radius $R$ is
\begin{equation}
    u v A_\perp \simeq \frac{GM_*\dot{M}}{R} = \frac{GM_*}{R} A_\perp \rho v,
\end{equation}
that implies virial scaling $u \propto p \propto \frac{GM}{R}\rho$. 
At the same time, for radiation-pressure-dominated matter, $u\propto \rho^{4/3}$.
Radiation energy density should thus depend on radius as $u\propto R^{-4}$, while $\pmag\propto R^{-6}$.
This is a general result implying that, whenever the cooling of an accretion column becomes inefficient, force-free condition is violated at a considerable elevation above its bottom.

One can see from Tables~\ref{tab:mod} and \ref{tab:res} that the condition for $\beta_{\rm BS}$ may be used as an applicability criterion for the analytic model: all the models with $\beta_{\rm BS} < 2/3$ are in good agreement with the analytic model, while large values of $\beta_{\rm BS}$ lead to column overheating and mass loss from the column at a finite height above the NS surface.
This is a geometry-dependent limit for the mass accretion rate, above which the analytic solution of \citetalias{BS76} predicts violation of the force-free condition above the surface of the NS.
The consequences of this violation will be discussed in Section~\ref{sec:disc:loss}.
On the other hand, as long as the position of the shock is much lower than the size of the magnetosphere, the analytic solution by \citetalias{BS76} gives a good approximation for the position of the shock even when the vent is at a finite height above the surface.

\subsection{Contribution of different physical and numerical effects}\label{sec:res:eff}

\subsubsection{Numerical resolution and photon diffusion}\label{sss.resolution}

\begin{figure}
\includegraphics[width=0.9\columnwidth]{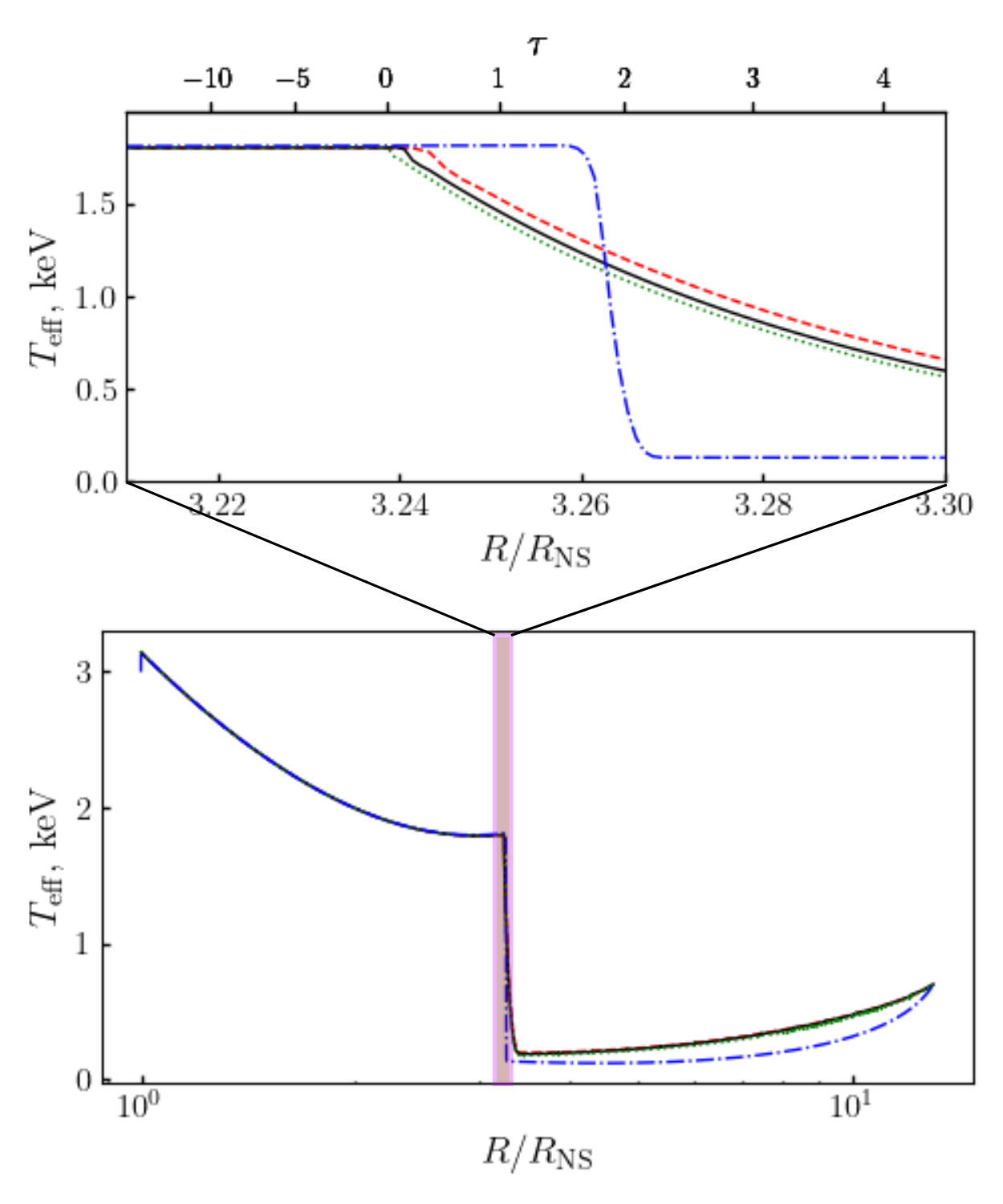}
 \caption{
 Effective temperature as a function of radius in models {\tt F} (black solid curve), {\tt L} (red dashed), {\tt F2} (green dotted), and {\tt ND} (blue dot-dashed).
 Upper panel shows a zoom-in into the shock region. 
 The secondary $x$ axis of the upper panel shows the optical depth along the field line (calculated for the model {\tt F2}), with the zero point at the shock front position. 
  }\label{fig:reseffects}       
\end{figure}

In Fig.~\ref{fig:reseffects}, we show how the distribution of effective temperature (calculated using the local radiation flux as $Q^- = \sigma_{\rm SB}T^4_{\rm eff}$) depends on spatial resolution: models {\tt F} (9600 cells), {\tt F2} (double resolution, 19200 cells), and {\tt L} (two times coarser resolution, 4800 cells) are shown together with the model {\tt ND} where photon diffusion along the field lines is turned off.
As we can see in the plot, the overall structure remains practically independent of resolution, which argues for the stability and reliability of the algorithm. 

Taking into account photon diffusion affects the shock position only slightly, by an amount comparable to the mean free path of a photon.
The profiles of all the physical quantities close to the shock front clearly change. 
The flow structure away from the shock, however, remains practically unaffected. 
In general, photon diffusion is most important on the spacial scales of the photon mean free path ($\sim 1/\varkappa \rho$). 
In Fig.~\ref{fig:reseffects}, we show the optical depth along the field line $\tau = \int \varkappa \rho {\rm d} l$ as a secondary axis. 
As we can see, the main effect of photon diffusion is the penetration of some of the energy into the upstream region, up to the depths of several. 
This agrees qualitatively with the picture drawn by the theory of  radiation-mediated shock waves in the diffusion approximation \citep{zeldovich-raizer, LN20}.

Upstream of the shock, the matter is radiating away the heat provided by the outer boundary condition (see Section~\ref{sec:num:BC}), with a small contribution of the photons diffusing from the post-shock region.
The effective temperature distribution in the optically thick part of the flow (see Fig.~\ref{fig:reseffects}) suggests a thermal spectrum with a characteristic temperature of $T \sim 1\div 3$~keV. 
However, a rigorous calculation of the observed spectrum should involve a complex combination of visibility conditions (different viewing angles and self-occultations for different parts of the flow). 
Besides, radiation coming from a NS accretion column should be strongly Comptonised. 
This Comptonisation may be saturated and non-saturated in different parts of the flow (see for example, \citealt{West17b}). 
Modelling the observational properties of accretion columns is left to a separate paper. 

For the fiducial model {\tt F}, the transverse optical depth in the immediate upstream region is about several. 
The optical depth of a single resolution element along the field line is in most of the models smaller than unity. 
At such spatial scales, the diffusion approximation is no longer valid, as photons can cross multiple resolution elements freely between scatterings. 
More elaborate description of the radiation fields near the shock front would reveal an additional narrow `Zeldovich spike' in energy density and temperature (see \citealt{tolstov15,fukue19} and references therein, and also \citealt{zeldovich-raizer}, Chapter VII). 

\subsubsection{Irradiation and rotation}

In models {\tt I}, {\tt WI}, and {\tt WI1}, irradiation is included in  momentum and energy equations as a force opposing gravity (see equations~\ref{E:src:s}, \ref{E:g_parallel}, and \ref{E:src:e}). 
We neglect the heating effect of irradiation.
Physically, this corresponds to the case when all the irradiating energy flux is immediately isotropically scattered. 
Self-irradiation in the form adopted in our paper is not fully self-consistent: the irradiating source is assumed to be located in the origin and radiate isotropically, though its luminosity is calculated by integrating the radiative losses from the simulated flow. 
Also, the radiation scattered from the flow is not included in the irradiating luminosity.
As long as most of the radiation losses occur near the NS surface, this is an accurate approximation in the outer magnetosphere. 

\begin{figure}
\includegraphics[width=0.9\columnwidth]{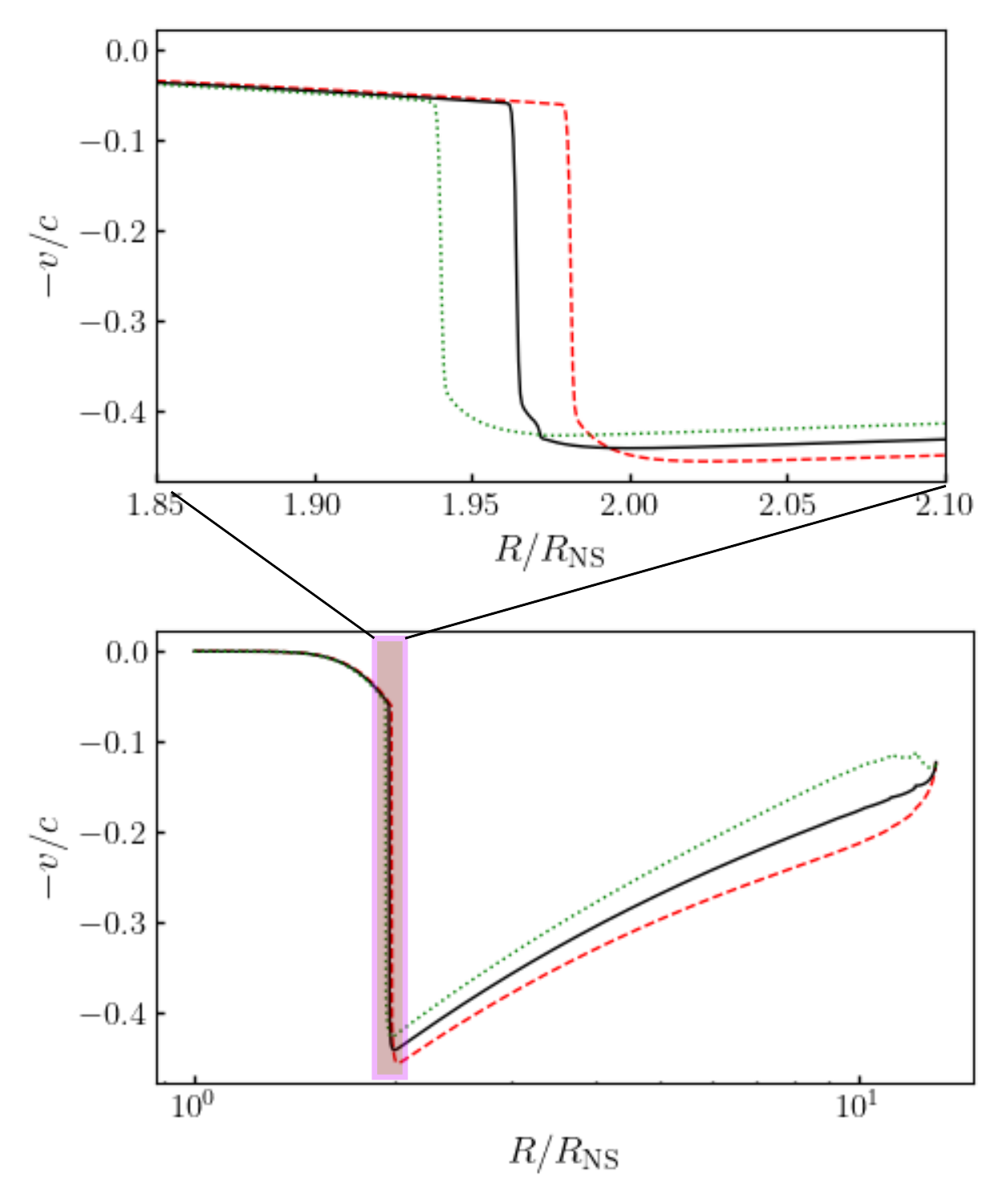}
 \caption{Time-averaged velocity dependence on radius for the models {\tt W} (red dashed curve) , {\tt WI} (black solid), and {\tt WI1} (green dotted). Averaging was done over the last 10 per cent of the simulation time.
As in Fig.~\ref{fig:reseffects}, the lower panel shows the entire range of radii, and the upper panel zooms into the shock front region.
 }\label{fig:irradeffects}       
\end{figure}

Irradiation mostly influences the outer parts of the flow, decreasing the infall velocity. 
This decreases the energy release in the column and thus, somewhat counter-intuitively, the height of the column (Fig.~\ref{fig:irradeffects}). 
For $\eta_{\rm irr} = 0.5$, the shock shifts by less than one per cent.
Increasing $\eta_{\rm irr}$ to the value of unity in {\tt WI1}, in addition to the overall shift of the shock front, makes the position of the shock unstable and excites oscillations on short time scales.
Probably, the nature of the oscillations is the same as the damped oscillations seen in all the models during their approach to equilibrium.
However, irradiation adds an additional positive feedback that compensates the dissipation of the oscillations. 
The shock position given in Table~\ref{tab:res} for {\tt WI1} is a time-averaged value, and the uncertainty here is dominated by the motion of the shock front rather than by the size of the spatial resolution element, as in the fiducial model. 

In models {\tt R} and {\tt RI}, we took into account centrifugal potential, tuned to match $0.9$ Keplerian rotation rate at $R_{\rm e}$. 
The main effect is the decrease in power released below the shock. 
Adding irradiation to rotation, surprisingly, leads to a larger luminosity and higher shock front position, which is probably related to the amount of energy lost in the sink. 

\subsection{Variability}\label{sec:res:var}

An interesting feature of practically all the simulations is their oscillatory behaviour during the shock settling stage. 
In all the simulations with radiation diffusion included, the position of the shock shows damped oscillations with a frequency close to the dynamical frequency at the relevant height. 
For example, for the fiducial model {\tt F}, oscillations are present while the shock front position changes in the range $R_{\rm shock} \sim 2-5~R_\NS$, \ver3{ where the dynamical} frequency is $f_{\rm dyn} = 1/t_{\rm dyn} \sim 200-800$~Hz. 

\begin{figure*}
\includegraphics[width=0.85\textwidth]{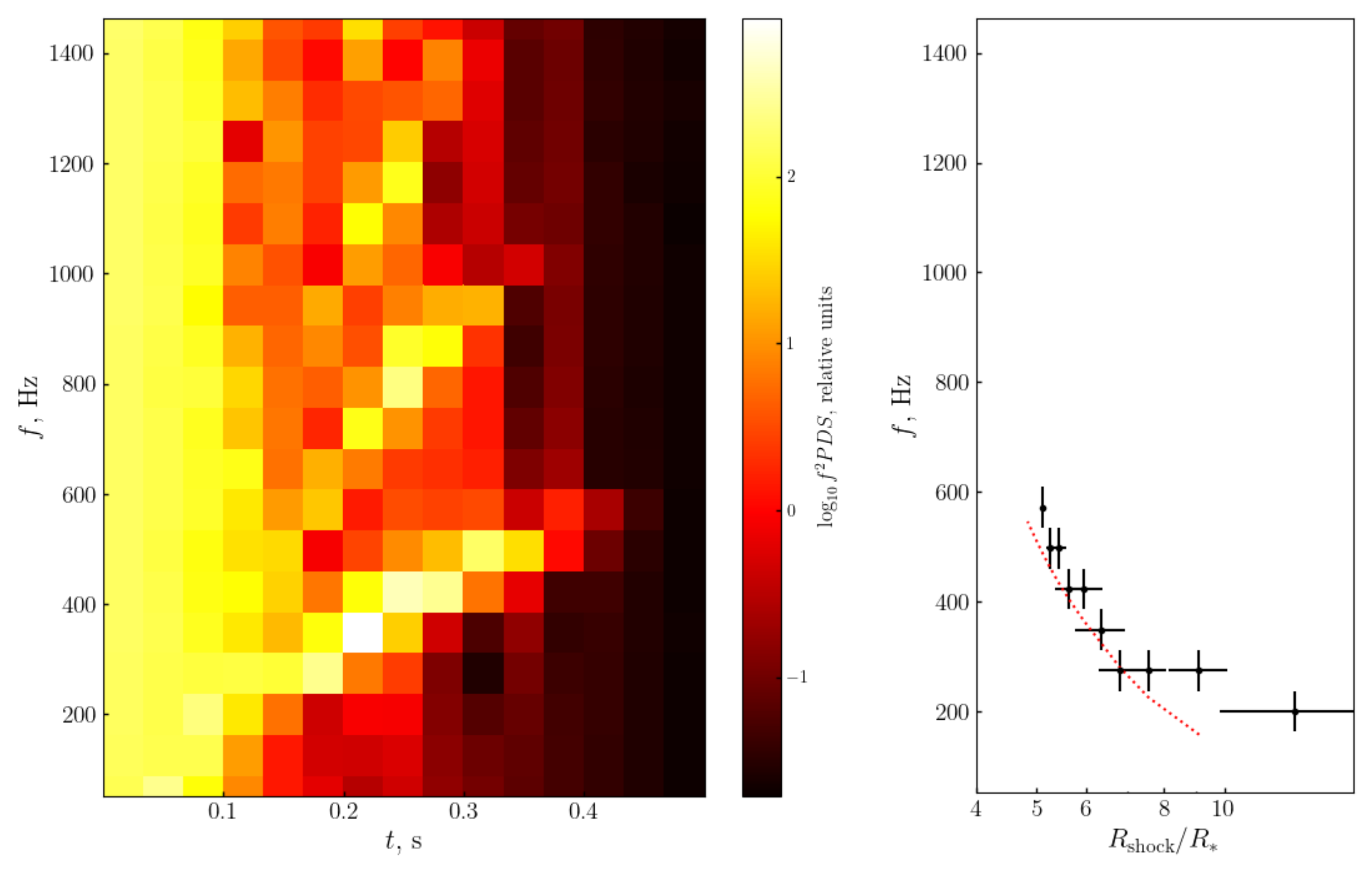}
 \caption{
    Left panel: dynamic power-density spectrum ($f^2 PDS$) for 15 uniform-length time intervals between $t=0$ and $0.5$\,s, model {\tt H}. 
    Right panel:  peak frequency for the same time bins as a function of shock position. 
    Sound-travel frequency $f_{\rm s}$ is shown in the right plot with the red dotted line.}\label{fig:dynplots}
\end{figure*}

Simulation {\tt H} has a larger magnetosphere size and hence a larger difference between the dynamical and replenishment time scales, allowing us to track the evolution of the oscillations during the settling of the shock front. 
In Fig.~\ref{fig:dynplots}, we show a temporally resolved power-density spectrum (PDS) for the total luminosity (calculated according to equation~\ref{E:ltot}) during the first $0.5$\,s of evolution. 
For each time bin, the PDS is calculated as the square of the absolute value of the Fourier image of $L_{\rm tot}$, normalised using Miyamoto normalisation \citep{miyamoto91,nowak99}
\begin{equation}
    PDS = 2\left| \frac{\int L(t)\, {\rm e}^{-2\uppi \i ft}\diff t}{\int L(t)\, \diff t}\right|^2,
\end{equation}
where $f$ is linear frequency.
For the larger portion of this time span, the radiation is highly variable at hectohertz frequencies. 
Most of the variability power is associated with the general settling trend, having a red spectrum with $PDS \propto f^{-2}$. 
Hence, in Fig.~\ref{fig:dynplots}, we show $f^2 PDS$, and the position of the peak in the spectrum is found as the maximum of this combination. 
This maximum evidently corresponds to a damped oscillation mode with an amplitude of about several per cent and a frequency changing with time from about 200 to 500$-$600\,Hz, as the shock moves downwards.

The observed peak frequencies may be reproduced by calculating the time needed for a sound wave to propagate from the top of the column to its base and back. 
Let us consider an accretion column in dynamic equilibrium, with a structure identical to the solution of \citetalias{BS76} in everything save the lower boundary condition (as the energy density has not necessarily reached the breakdown limit). 
The time needed to travel between $R_*$ and $R_{\rm shock}$ is, according to equations~(\ref{E:geo:l}) and (\ref{E:time:sound}),
\begin{equation}
    \Delta t_{\rm s} = \int_{\rm R_*}^{R_{\rm shock}} \frac{\sqrt{3\cos^2\theta+1}}{2\cos\theta} \frac{\,{\rm d}R}{c_{\rm s}},
\end{equation}
where
\begin{equation}
\displaystyle    c_{\rm s} \simeq \sqrt{\frac{4}{9}\frac{u}{\rho}} \simeq \sqrt{\frac{1}{3} \frac{GM_*}{R_*} e^{\gamma \xi} \left( \frac{1}{\xi} \eint{2}{\gamma\xi} + \beta_{\rm BS} e^{-\gamma} - \eint{2}{\gamma}\right)},
\end{equation}
and $\xi = R/R_*$. 
While $\gamma$ depends only on geometry, $R_{\rm shock}$ and $\beta_{\rm BS}$ vary with time as they depend on the value of $u$ at the NS surface.
In the right panel of Fig.~\ref{fig:dynplots}, the corresponding frequency $f_{\rm s} = 1/\Delta t_{\rm s}$ is shown for comparison. 
As $\Delta t_{\rm s}$ is the time needed for the sound to propagate only in one direction, from the shock front to the surface (or back), the oscillation mode we see in the simulations is likely the first overtone. 

Apart from the main oscillation frequency, gradually changing with time, it is possible to detect its first two harmonics, with the frequencies two and three times larger (see the left panel of Fig.~\ref{fig:dynplots}). 
Their amplitudes are at least one order of magnitude smaller, and they are less likely to be detectable in real astrophysical sources.
Damping of the oscillations is likely physical rather than numerical, as its rate is about the same for different resolutions. 
It is also possible that the oscillations are in fact powered by the evolving structure of the accretion column.

In most of the simulations, oscillations disappear after several replenishment times. 
The only exceptions are models {\tt N2} and {\tt M100}, where the analytic solution is self-inconsistent, and model {\tt WI1}, where irradiation apparently supports the oscillation process.  
In {\tt N2} and {\tt M100}, there is also mass loss variation with the oscillation phase, by about 20 per cent and by more than an order of magnitude, respectively. 

\section{Discussion}\label{sec:disc}

\subsection{Matter lost from the column}\label{sec:disc:loss}

Introducing the mass-loss law (equation~\ref{E:src:m}), we assume
that, whenever the thermal pressure significantly exceeds the pressure of the magnetic field, mass is lost from the column at about the speed of sound. 
This leakage carries away also internal energy and momentum.
This is a simplified picture, but it helps justify the inner boundary condition and extend the approach of \citetalias{BS76} to the parameter range where, their solutions become internally inconsistent~(see section~\ref{sec:res:breakdown}).

For moderate mass accretion rates (more precisely, for moderate advection parameters $\beta_{\rm BS} < 2/3$), accretion column tends to leak at the bottom. 
The physics behind this process likely involves three-dimensional interchange instabilities \citep{Arons-Lea1976} that start working when thermal pressure starts exceeding that of the magnetic field or a critical value scaling with the magnetic field pressure \citep{litwin01,mukherjee13,Kulsrud-Sunyaev}.
The matter leaving the column ultimately cools down and spreads over the surface of the NS \citep{MP01}. 

Large advection parameters cause the column to leak at a finite height. 
If a small amount of plasma leaves the column at a considerable height, it becomes immersed in a strong, super-Eddington, anisotropic external radiation field.
Depending on the structure of the radiation field, the matter leaving the column may be either accreted or expelled along the field lines. 
We expect the realistic picture to be a combination of the three possible outcomes: (i) the cross-section
of the column increases closer to the surface of the star, as the matter is spread over adjacent field lines; (ii) plasma is lost from the column onto closed magnetic field lines, and remains trapped inside the magnetosphere; (iii) plasma is ejected along the open field lines, acquiring mildly relativistic velocities.

If all the matter remains gravitationally bound but spreads onto adjacent field lines, the column may be considered as having larger cross-section and perimeter. 
Increasing $\delta$ leads to larger $\beta_{\rm BS}$, while increasing azimuthal size decreases the effective value of $\beta_{\rm BS}$. 
Depending on the initial shape of the magnetospheric flow and the instability increments in different parts of the perimeter, effective $\beta_{\rm BS}$ may change differently. 
If both the latitudinal and the longitudinal dimensions of the column increase by the same small amount of $h$, and $A_\perp /\delta^2 \gg 1$, both thickness $\delta$ and cross-section $A_\perp$ increase linearly with $h$.
This implies decreasing ratio $A_\perp / \delta^2$, smaller $\gamma$ (equation~\ref{E:BS:gamma}), and larger $\beta_{\rm BS}$ (equation~\ref{E:BS:beta}).
Isotropic expansion of the column in general leads to smaller perimeter to cross-section ratio and, consequently, complicates cooling. 

If the column has full azimuthal coverage ($a=1$), the cross-section grows without any changes in the perimeter length. 
This is well illustrated by the sequence of models {\tt M100W3-50}, that differ only in the latitudinal extent of the flow, set by the penetration depth at the outer boundary $\Delta R_{\rm e}$. 
Here, larger $\delta(R_\NS)$ leads to larger advection parameters. 
As longitudinal spreading of the column decreases the perimeter to cross-section ratio, it hinders the cooling of the column, leading to even larger effective values of $\beta_{\rm BS}$ and more spreading. 
One might check that increasing $\Delta R_{\rm e}$ in this case always leads to a decrease in $\gamma$ and an increase in $\xi_{\rm s}$. 
According to equation~(\ref{E:BS:beta}), this leads to larger values of $\beta_{\rm BS}$.

The mass lost to the closed lines is likely to be trapped in certain regions of the magnetosphere and accumulate until its optical depth becomes larger than unity. 
Large optical depth decreases the contribution of the radiation pressure force and allows the matter to fall onto the star. 
The mass trapped on closed field lines is potentially a source of stochastic or quasi-periodic variability on the dynamical time scales of the outer magnetosphere \citep{AB20}. 

The mass lost along the open field lines will likely acquire mildly relativistic velocities weakly dependent on the Eddington factor. 
If the Eddington limit is strongly violated, acceleration of the optically thin material is 
\begin{equation}
    \frac{{\rm d}v}{{\rm d}t} \simeq \frac{\varkappa L }{4\uppi cR^2}.
\end{equation}
After integration along the approximately radial trajectory starting from $R\simeq R_*$, this results in
\begin{equation}
    v_{\rm out} \simeq \sqrt{\frac{\varkappa L}{4\pi c R_*}} = \sqrt{\frac{L}{L_{\rm Edd}}\frac{GM_*}{R_*}} \simeq 0.4 \sqrt{\frac{L}{L_{\rm Edd}}\frac{M_*}{1.4\Msun} \frac{11\,\rm km}{R_*}}\, c.
\end{equation}
This provides XRPs with a mechanism to launch mildly relativistic outflows. 
If the outflow velocity becomes relativistic, the efficiency of the radiation pressure acceleration rapidly decreases, and Compton drag becomes important, making it essentially impossible to accelerate a flow by radiation pressure to velocities $\gtrsim 0.5c$ \citep{phinney82}. 
Besides, non-negligible optical depth of the flow itself will decrease the effective luminosity value.

The electron-scattering optical depth of such an outflow (if not strongly collimated) is 
\begin{equation}
    \tau_{\rm out} \sim \int_{R_*}^{+\infty} \frac{\varkappa \dot{M}_{\rm out}}{4\pi R^2 v(R)}\,{\rm d}R \simeq \frac{c}{v_{\rm out}} \frac{\dot{M}_{\rm out}c^2}{L_{\rm Edd}} \gtrsim 1,
\end{equation}
where $\dot{M}_{\rm out}$ is the mass-loss rate. 
Favourable (face-on) orientation of the source will allow one to see the X-ray radiation of the central source through the outflowing matter.
In this case, the observer will see blueshifted absorption features similar to those observed in some ultraluminous X-ray sources \citep{Pinto17,Kosec+2018,pinto21}.

Thus, both black hole and NS accretors are likely to launch outflows with the velocities of the order of tenths of the speed of light, as this is the characteristic virial velocity in both cases. 
In particular, such outflows may be formed in the magnetospheres of highly magnetised NSs. 
Existence of a mildly relativistic outflow does not necessarily imply an accretion disc close to the compact object. 

If, for some reasons, the mass loss from the column does not start at all, one would expect the cross-section of the column to increase due to deformation of the field lines.
For the column to reach a steady state, this deformation should facilitate the cooling of the column. 
As we have seen earlier this section, this is possible only if the perimeter of the column grows faster than its cross-section. 
For a geometrically thin column with $A_\perp / \delta^2 \gg 1$, this is unlikely, as roughly uniform deformation of the field lines is expected to make the cross-section rounder.
While it is possible that some process might effectively increase the radiating surface of the column, it would probably involve extensive field line deformation inevitably leading to reconnection. 
Reconnection would imply mass spreading onto adjacent field lines and hence mass loss from the column. 

\subsection{High-frequency oscillations of accretion columns}\label{sec:disc:osc}

Our simulations clearly show an oscillation mode at the sound propagation time scale of the accretion column (see section~\ref{sec:res:var}). 
At the same time, in the observational data for real X-ray pulsars, there is little evidence for high-frequency ($\sim 100-1000$Hz) variability. 
The timing behaviour of the classic X-ray pulsars (long-spin-period accreting NSs in HMXB) is usually consistent with the assumption of all the variability coming from the accretion disc and maybe the outermost parts of the magnetosphere \citep{revnivtsev09,RN13}.
The frequency range above several tens of hertz is usually not covered due to poor statistics and interference from dead-time effects. 
In particular, for {\it RXTE/PCA}, dead time becomes an important factor at frequencies about 1kHz \citep{jahoda06,revnivtsev15}. 

The `bursting pulsar' \object{GRO J1744$-$28} \citep{dai15} shows a broad-band peaked noise component at hectohertz frequencies. 
In \citet{klein96}, this variability component is interpreted as a manifestation of the photon bubble instability \citep{arons92} in the accretion column of the pulsar.
Given the uniqueness of this object, it is hardly representative of the population of XRPs in general. 

Probably, the only example of a conventional XRP with a reported detection of high-frequency variability is \object{Cen~X-3}. 
It is a high-mass X-ray binary consisting of an OB donor star \citep{RJ83} and a NS with the magnetic field of $(2\div 3)\times 10^{12}\,$G (measured using a cyclotron line, see \citealt{santangelo98}). 
The deep timing study performed by \citet{jernigan00} revealed a broad spectral feature at about $1$kHz.
While in the original paper this high-frequency variability was interpreted as a manifestation of the photon bubble instability, its properties are also consistent with the global oscillations of an accretion column. 
The luminosity ($L \sim 10^{38}\ergl$) and the magnetic field of the object (magnetic moment $\mu \sim 3\times 10^{30}{\rm \, G\, cm^{3}}$) suggest the shock is located at  $R_{\rm shock} \sim 2R_\NS$, and the sound propagation time is $10^{-4}-10^{-2}\,$s, depending on the details of the geometry. 
However, the results reported by \citet{jernigan00} are
questionable because of the interfering instrumental effects, that are likely to produce excess power around $\sim 1\,$kHz. 
The possible timing artefacts are considered by \citet{revnivtsev15}, who show that the spectral shapes of XRPs are unfavourable for timing studies at high frequencies, at least using \textit{RXTE} data. 

\ver3{Modern instruments like {\it HMXT/Insight} \citep{Li+2007} are capable of timing analysis at frequencies up to $\sim 10$~kHz.
During the 2017-2018 outburst of \object{Swift~J0243.6+6124}, the object was extensively studied by {\it HMXT/Insight} \citep{Doroshenko+2020}, and no variability features above 100~Hz were detected (V.~Doroshenko, private communication).}

In accreting millisecond X-ray pulsars (AMXP, see \citealt{AMXP_review} for review), there is significant variability at the time scales comparable to the dynamical time scales near the surface of the accreting NS \citep{vstraaten05}. 
For instance, they are known to have quasi-periodic oscillations (QPOs) in the hecto- and kilohertz range, similar to those observed in atoll sources \citep{WK99,vstraaten05}. 
 However, luminosities of AMXPs rarely exceed $10^{37}\ergl$, which practically excludes the formation of  a radiation-dominated accretion column.
AMXPs are known to have compact magnetospheres ($R_{\rm e}/R_\NS \lesssim 5$). 
The critical luminosity required for column formation $L^*$, according to the equation (1) of \citetalias{BS76} and formulae from our Appendix~\ref{app:geo}, may be estimated for such objects as
 \begin{equation}
 \label{E:Lcrit}
 \displaystyle
L^* \sim  a \, \sqrt{\frac{R_*}{R_{\rm e}} }   \, L_{\rm Edd}\, \sim 10^{38}\ergl.
 \end{equation}
Rapid variability of AMXPs, in many aspects similar to the variability of non-magnetised NSs, is probably related to the inner disc or boundary layer rather than to a magnetic column \citep{vdK_review}. 

The QPO seen in our simulation {\tt H} has a large quality factor (probably larger than 10, see Fig.~\ref{fig:dynplots}), and its frequency is a strong function of the luminosity of the column. 
This is a direct consequence of the one-dimensional nature of our simulations: all the parameters depend only on a single spatial coordinate, implying a single sound propagation time between the shock wave and the surface of the star.
A realistic, multi-dimensional accretion column is unlikely to have a single shock front position for all the field lines, and the sound propagation times along different field lines should differ accordingly.
Hence we suggest that, instead of a narrow QPO peak, a realistic accretion column should have a broad-band peaked noise excess in its PDS. 

Qualitatively similar oscillations of the shock front were found in numerical simulations of accretion columns in T~Tau stars \citep{koldoba08} and cataclysmic variables \citep{Bera18}. 
The physical conditions in both cases are, however, different from the accretion columns in NS systems. 
In particular, accretion columns formed during accretion onto young stars and white dwarfs are usually optically thin. 
Thus, radiation pressure is usually a minor factor. 
The oscillations found by \citet{Bera18}, involving weak secondary shocks propagating between the surface of the accretor and the primary shock front, also have a period well approximated by the sound crossing time scale. 
Interestingly, certain magnetic cataclysmic variables indeed have oscillation modes at about 1\,Hz, suggestive for the column oscillation times. A comprehensive list of magnetic CVs showing QPOs with frequencies $\sim 1$\,Hz is given by \citet{BB15}. 

In \citet{Zhang+2021,zhang22,zhang23}, detailed numerical models of NS accretion columns were developed using two-dimensional radiative MHD modelling.
These studies focus on the steady-state picture and at the same time do not have a mass sink neither as a source term nor as a boundary condition, that makes the result relevant for the mass accumulation stage. 
The authors have found quasi-periodic oscillations with frequencies between 2 and 25kHz. 
The oscillation mode is interpreted as quasi-hydrostatic heating/cooling cycle driven by heat accumulation inside the column.
Their results are difficult to compare to ours because of the differences in boundary conditions (in particular, \citealt{zhang22,zhang23} assume very efficient cooling at the bottom) and the simulation setup in general. 

\subsection{Observability of the replenishment time scale}

As we have seen in section~\ref{sec:times}, column replenishment time scales in most XRPs should be longer than the dynamical times at the magnetosphere boundary.
At the same time, most of the energy emitted by a rapidly accreting XRP normally comes from the accretion column.
Does this mean that all the variability of the disc at the time scales shorter than the replenishment time will be damped? 
This is an issue we are planning to address in a separate paper. 

It is well known that the power density spectra of X-ray pulsars have a break at a frequency close to the spin frequency of the NS. 
The position of the break changes with time and apparently correlates with the flux \citep{revnivtsev09}.
If the break frequency corresponds to the dynamical time scale at the radius of the magnetosphere, we expect $f_{\rm break} \propto \dot{M}^{3/7}$. 
However, if the spectrum is cut off at the replenishment time, we expect a different scaling $f_{\rm break} \propto \dot{M}^{5/7} (\Delta R_{\rm e}/R_{\rm e})^{-1}$, involving the variations of the penetration depth in addition to the magnetosphere radius.
For typical X-ray pulsar parameters, we also expect the values of the break frequency to be smaller if related to the replenishment time, see Eq.~\eqref{E:trep:RA}.

An important observational test for replenishment time would be to track the transition from the accretor to the propeller stage. 
However, this transition should occur at a luminosity large enough to support an accretion column, presumably above $\sim 10^{36}-10^{37}\ergl$ (equation~\ref{E:Lcrit}). 
On the other hand, transition between accretor and propeller states occurs in a wide range of luminosities $L \sim 10^{35}-10^{37}\ergl$ \citep{Cui97, campana01, Tsygankov16}. 
Most relevant candidate events are the two turn-off transitions observed for \object{4U\,0115+63} and \object{V\,0332+53} \citep{Tsygankov16}. 
Observational data allow one to estimate the flux decrease e-folding time scales during these events as about 14.3 and 5.6 hours, respectively.

For an XRP with the spin period $P_{\rm s} = 2\uppi / \Omega$,  replenishment time scale  may be estimated by substituting $\xim R_{\rm A} = \left( GM/\Omega^2\right)^{1/3}$ to (\ref{E:trep}), in the assumption that the size of the magnetosphere coincides with the co-rotation radius $(GM/\Omega^2)^{1/3}$.  
Finally, we get
\begin{equation}
\displaystyle
\begin{array}{lcc}
     t_{\rm r} &=   &  \displaystyle (2\uppi)^{-2/3}\frac{a}{2^{1/6}} \frac{\Delta R_{\rm e}}{R_{\rm e}} \xim^{-8/3} \frac{\left(GM_*\right)^{1/3}}{R_*} P_{\rm spin}^{5/3} \\
 & \simeq  &   \displaystyle  180  \frac{a}{\xim^{8/3}} \frac{\Delta R_{\rm e}}{R_{\rm e}} \left( \frac{M_*}{1.5\Msun}\right)^{1/3} \frac{10\km}{R_*} \left( \frac{P_{\rm spin}}{1\rm s}\right)^{5/3} {\rm s}.
\end{array}
\end{equation}
Note the strong dependence on $\xim$ allows the actual replenishment time to vary in broad limits, from seconds to hours.
For the parameters of \object{4U\,0115+63} (spin period $P_{\rm spin} \simeq 3.62$s, magnetic moment $\mu \simeq 10^{30} {\rm G \, cm^3}$), the above expression may be used to explain a process taking several hours.
However, fast-decaying light curves are also expected to appear as a result of the evolution of a viscous accretion disc with a cooling front (see, for example, \citealt{lipunova22}). 

An important outcome of the long replenishment time scale is that realistic accretion columns should be out of equilibrium. 
Near the break frequency, the stochastic component in a PDS of an XRP has a power density of the order $0.01\,{\rm Hz^{-1}}$ (relevant frequencies are about $1$Hz; see, e. g., \citealt[figure 14]{reig11} or \citealt[figure 1]{revnivtsev09}), that suggests accretion rate variations at the level of $\sim 10\%$ at the dynamic time scales of the inner disc.
Disc variability on all the time scales less than the replenishment time of the column contributes to the variations of the mass accretion rate.
Therefore, the mass accretion rate might change within a single replenishment time by a factor considerably larger than the above estimate of $10\%$.

\section{Conclusions}\label{sec:conc}

Modelling the dynamics of a NS accretion column allows us to reproduce the main results of the analytic model of \citetalias{BS76}. 
In a broad range of parameter values, the analytic model accurately predicts the position of the shock. 
The structure of the sinking region below the shock is also well reproduced.

Analysing the consistency of the models at high mass accretion rates, we find a new and simple criterion that restricts the applicability of the analytic model: the advection parameter $\beta_{\rm BS}$, incorporating mass accretion rate and column geometry, should not exceed $2/3$. 
At large mass accretion rates and optical depths, when radiative cooling becomes inefficient, the vertical pressure profile becomes shallower than the vertical profile for magnetic pressure.
As a result, thermal pressure becomes dominant over magnetic pressure not at the base of the column, as it was assumed by \citetalias{BS76}, but at a finite height above the surface of the NS. 

This conjecture, following from analytic considerations, is also supported by our numerical results.
For the models with a large advected fraction of the accretion power (more than $2/3$), breakdown of the force-free condition and the associated mass loss tends to occur at a finite height, and agreement with the analytic solution becomes worse. 

Mass loss from a finite height should lead to spreading of the column (its transverse size becomes larger than expected from the dipole geometry of the problem) and/or mass-loading of the magnetic field lines.
Mass loading of the open field lines is a possible way to create mildly relativistic outflows from a magnetised accreting NS.

The characteristic time scale at which an accretion column evolves to the stationary solution may be described as the mass replenishment time. 
This is the time needed to accumulate enough mass to put the bottom of the column to the edge of a breakdown (thermal and magnetic pressures become equal).  
For a large magnetosphere, this time scale generally exceeds the dynamical time scale of the inner accretion disc, making it a potentially important feature in the variability spectra of XRPs. 

In particular,  finite lifetime of an accretion column may affect the behaviour during accretor-propeller transitions, if the transition occurs from the state with a radiation-supported accretion column.

We find that the approach to the equilibrium state is non-monotonic, showing damped oscillations of the shock front position around the average trend. 
The frequency of this oscillation mode corresponds to the sound propagation time between the shock wave and the NS surface. 
For actual XRPs, the frequency falls in the hecto- to kilohertz range.
\ver3{In most of our simulations, the oscillations are damped on the time scales of the order of the replenishment time or faster, that may explain their elusiveness in the observational data.}

\section*{Acknowledgements}

We acknowledge the support from the Academy of Finland grants 332666 and 333112. 
We also acknowledge funding from Russian Science Foundation grant 21-12-00141 and the usage of computing resources of the M. V. Lomonosov Moscow State University Program of Development for an analysis of the stability of the model.
The authors would like to thank Sergey Tsygankov \ver3{and Victor Doroshenko for discussions about timing of XRPs, Amir Levinson and Evgeny Derishev for discussions about radiative shocks,} and Indrek Vurm for a general discussion on accretion column modelling.
We acknowledge the use of {\tt MayaVI} 3D-plotting software \citep{mayavi}.

\section*{Data Availability}

The numerical code {\tt HACol} is freely available on \url{https://github.com/pabolmasov/HACol}. All the simulation results are available upon request. 



\bibliographystyle{mnras}
\bibliography{mybib} 

\hyphenation{Post-Script Sprin-ger}
\begin{thebibliography}{}
\makeatletter
\relax
\def\mn@urlcharsother{\let\do\@makeother \do\$\do\&\do\#\do\^\do\_\do\%\do\~}
\def\mn@doi{\begingroup\mn@urlcharsother \@ifnextchar [ {\mn@doi@}
  {\mn@doi@[]}}
\def\mn@doi@[#1]#2{\def\@tempa{#1}\ifx\@tempa\@empty \href
  {http://dx.doi.org/#2} {doi:#2}\else \href {http://dx.doi.org/#2} {#1}\fi
  \endgroup}
\def\mn@eprint#1#2{\mn@eprint@#1:#2::\@nil}
\def\mn@eprint@arXiv#1{\href {http://arxiv.org/abs/#1} {{\tt arXiv:#1}}}
\def\mn@eprint@dblp#1{\href {http://dblp.uni-trier.de/rec/bibtex/#1.xml}
  {dblp:#1}}
\def\mn@eprint@#1:#2:#3:#4\@nil{\def\@tempa {#1}\def\@tempb {#2}\def\@tempc
  {#3}\ifx \@tempc \@empty \let \@tempc \@tempb \let \@tempb \@tempa \fi \ifx
  \@tempb \@empty \def\@tempb {arXiv}\fi \@ifundefined
  {mn@eprint@\@tempb}{\@tempb:\@tempc}{\expandafter \expandafter \csname
  mn@eprint@\@tempb\endcsname \expandafter{\@tempc}}}

\bibitem[\protect\citeauthoryear{{Abarca}, {Parfrey}  \&
  {Klu{\'z}niak}}{{Abarca} et~al.}{2021}]{Abarca+2021}
{Abarca} D.,  {Parfrey} K.,   {Klu{\'z}niak} W.,  2021, \mn@doi [\apjl]
  {10.3847/2041-8213/ac1859}, \href
  {https://ui.adsabs.harvard.edu/abs/2021ApJ...917L..31A} {917, L31}

\bibitem[\protect\citeauthoryear{{Abolmasov} \& {Biryukov}}{{Abolmasov} \&
  {Biryukov}}{2020}]{AB20}
{Abolmasov} P.,  {Biryukov} A.,  2020, \mn@doi [\mnras]
  {10.1093/mnras/staa1544}, \href
  {https://ui.adsabs.harvard.edu/abs/2020MNRAS.496...13A} {496, 13}

\bibitem[\protect\citeauthoryear{{Arons}}{{Arons}}{1992}]{arons92}
{Arons} J.,  1992, \mn@doi [\apj] {10.1086/171174}, \href
  {https://ui.adsabs.harvard.edu/abs/1992ApJ...388..561A} {388, 561}

\bibitem[\protect\citeauthoryear{{Arons} \& {Lea}}{{Arons} \&
  {Lea}}{1976}]{Arons-Lea1976}
{Arons} J.,  {Lea} S.~M.,  1976, \mn@doi [\apj] {10.1086/154562}, \href
  {https://ui.adsabs.harvard.edu/abs/1976ApJ...207..914A} {207, 914}

\bibitem[\protect\citeauthoryear{{Arons}, {Klein}  \& {Lea}}{{Arons}
  et~al.}{1987}]{arons87}
{Arons} J.,  {Klein} R.~I.,   {Lea} S.~M.,  1987, \mn@doi [\apj]
  {10.1086/164912}, \href
  {https://ui.adsabs.harvard.edu/abs/1987ApJ...312..666A} {312, 666}

\bibitem[\protect\citeauthoryear{{Bachetti}, {Romanova}, {Kulkarni}, {Burderi}
  \& {di Salvo}}{{Bachetti} et~al.}{2010}]{bachetti10}
{Bachetti} M.,  {Romanova} M.~M.,  {Kulkarni} A.,  {Burderi} L.,   {di Salvo}
  T.,  2010, \mn@doi [\mnras] {10.1111/j.1365-2966.2010.16203.x}, \href
  {https://ui.adsabs.harvard.edu/abs/2010MNRAS.403.1193B} {403, 1193}

\bibitem[\protect\citeauthoryear{{Bachetti} et~al.,}{{Bachetti}
  et~al.}{2014}]{bachetti14}
{Bachetti} M.,  et~al., 2014, \mn@doi [\nat] {10.1038/nature13791}, \href
  {https://ui.adsabs.harvard.edu/abs/2014Natur.514..202B} {514, 202}

\bibitem[\protect\citeauthoryear{{Basko} \& {Sunyaev}}{{Basko} \&
  {Sunyaev}}{1975}]{BS75}
{Basko} M.~M.,  {Sunyaev} R.~A.,  1975, \aap, \href
  {https://ui.adsabs.harvard.edu/abs/1975A&A....42..311B} {42, 311}

\bibitem[\protect\citeauthoryear{{Basko} \& {Sunyaev}}{{Basko} \&
  {Sunyaev}}{1976}]{BS76}
{Basko} M.~M.,  {Sunyaev} R.~A.,  1976, \mn@doi [\mnras]
  {10.1093/mnras/175.2.395}, \href
  {http://adsabs.harvard.edu/abs/1976MNRAS.175..395B} {175, 395}

\bibitem[\protect\citeauthoryear{{Becker}}{{Becker}}{1998}]{Becker1998}
{Becker} P.~A.,  1998, \mn@doi [\apj] {10.1086/305568}, \href
  {https://ui.adsabs.harvard.edu/abs/1998ApJ...498..790B} {498, 790}

\bibitem[\protect\citeauthoryear{{Becker} \& {Wolff}}{{Becker} \&
  {Wolff}}{2007}]{BW07}
{Becker} P.~A.,  {Wolff} M.~T.,  2007, \mn@doi [\apj] {10.1086/509108}, \href
  {https://ui.adsabs.harvard.edu/abs/2007ApJ...654..435B} {654, 435}

\bibitem[\protect\citeauthoryear{{Bera} \& {Bhattacharya}}{{Bera} \&
  {Bhattacharya}}{2018}]{Bera18}
{Bera} P.,  {Bhattacharya} D.,  2018, \mn@doi [\mnras] {10.1093/mnras/stx2720},
  \href {https://ui.adsabs.harvard.edu/abs/2018MNRAS.474.1629B} {474, 1629}

\bibitem[\protect\citeauthoryear{{Bernstein}, {Frieman}, {Kruskal}  \&
  {Kulsrud}}{{Bernstein} et~al.}{1958}]{Bernstein+1958}
{Bernstein} I.~B.,  {Frieman} E.~A.,  {Kruskal} M.~D.,   {Kulsrud} R.~M.,
  1958, \mn@doi [Proceedings of the Royal Society of London Series A]
  {10.1098/rspa.1958.0023}, \href
  {https://ui.adsabs.harvard.edu/abs/1958RSPSA.244...17B} {244, 17}

\bibitem[\protect\citeauthoryear{{Bildsten} et~al.,}{{Bildsten}
  et~al.}{1997}]{bildsten97}
{Bildsten} L.,  et~al., 1997, \mn@doi [\apjs] {10.1086/313060}, \href
  {https://ui.adsabs.harvard.edu/abs/1997ApJS..113..367B} {113, 367}

\bibitem[\protect\citeauthoryear{{Bonnet-Bidaud}, {Mouchet}, {Busschaert},
  {Falize}  \& {Michaut}}{{Bonnet-Bidaud} et~al.}{2015}]{BB15}
{Bonnet-Bidaud} J.~M.,  {Mouchet} M.,  {Busschaert} C.,  {Falize} E.,
  {Michaut} C.,  2015, \mn@doi [\aap] {10.1051/0004-6361/201425482}, \href
  {https://ui.adsabs.harvard.edu/abs/2015A&A...579A..24B} {579, A24}

\bibitem[\protect\citeauthoryear{{Brumback}, {Hickox}, {F{\"u}rst},
  {Pottschmidt}, {Tomsick}, {Wilms}, {Staubert}  \& {Vrtilek}}{{Brumback}
  et~al.}{2021}]{herX1}
{Brumback} M.~C.,  {Hickox} R.~C.,  {F{\"u}rst} F.~S.,  {Pottschmidt} K.,
  {Tomsick} J.~A.,  {Wilms} J.,  {Staubert} R.,   {Vrtilek} S.,  2021, \mn@doi
  [\apj] {10.3847/1538-4357/abe122}, \href
  {https://ui.adsabs.harvard.edu/abs/2021ApJ...909..186B} {909, 186}

\bibitem[\protect\citeauthoryear{{Caballero} \& {Wilms}}{{Caballero} \&
  {Wilms}}{2012}]{XRPreview}
{Caballero} I.,  {Wilms} J.,  2012, \memsai, \href
  {https://ui.adsabs.harvard.edu/abs/2012MmSAI..83..230C} {83, 230}

\bibitem[\protect\citeauthoryear{{Campana}, {Gastaldello}, {Stella}, {Israel},
  {Colpi}, {Pizzolato}, {Orland ini}  \& {Dal Fiume}}{{Campana}
  et~al.}{2001}]{campana01}
{Campana} S.,  {Gastaldello} F.,  {Stella} L.,  {Israel} G.~L.,  {Colpi} M.,
  {Pizzolato} F.,  {Orland ini} M.,   {Dal Fiume} D.,  2001, \mn@doi [\apj]
  {10.1086/323317}, \href
  {https://ui.adsabs.harvard.edu/abs/2001ApJ...561..924C} {561, 924}

\bibitem[\protect\citeauthoryear{{Chandrasekhar}}{{Chandrasekhar}}{1967}]{chandra67}
{Chandrasekhar} S.,  1967, {An introduction to the study of stellar structure}

\bibitem[\protect\citeauthoryear{{Courant}, {Friedrichs}  \& {Lewy}}{{Courant}
  et~al.}{1967}]{CFL}
{Courant} R.,  {Friedrichs} K.,   {Lewy} H.,  1967, \mn@doi [IBM Journal of
  Research and Development] {10.1147/rd.112.0215}, \href
  {http://adsabs.harvard.edu/abs/1967IBMJ...11..215C} {11, 215}

\bibitem[\protect\citeauthoryear{{Cui}}{{Cui}}{1997}]{Cui97}
{Cui} W.,  1997, \mn@doi [\apjl] {10.1086/310712}, \href
  {https://ui.adsabs.harvard.edu/abs/1997ApJ...482L.163C} {482, L163}

\bibitem[\protect\citeauthoryear{{D'A\`\i} et~al.,}{{D'A\`\i}
  et~al.}{2015}]{dai15}
{D'A\`\i} A.,  et~al., 2015, \mn@doi [\mnras] {10.1093/mnras/stv531}, \href
  {https://ui.adsabs.harvard.edu/abs/2015MNRAS.449.4288D} {449, 4288}

\bibitem[\protect\citeauthoryear{{Davidson}}{{Davidson}}{1973}]{Davidson1973}
{Davidson} K.,  1973, \mn@doi [Nature Physical Science]
  {10.1038/physci246001a0}, \href
  {https://ui.adsabs.harvard.edu/abs/1973NPhS..246....1D} {246, 1}

\bibitem[\protect\citeauthoryear{{Doroshenko} et~al.,}{{Doroshenko}
  et~al.}{2020}]{Doroshenko+2020}
{Doroshenko} V.,  et~al., 2020, \mn@doi [\mnras] {10.1093/mnras/stz2879}, \href
  {https://ui.adsabs.harvard.edu/abs/2020MNRAS.491.1857D} {491, 1857}

\bibitem[\protect\citeauthoryear{{Einfeldt}}{{Einfeldt}}{1988}]{HLLE_Einfeldt}
{Einfeldt} B.,  1988, \mn@doi [SIAM Journal on Numerical Analysis]
  {10.1137/0725021}, \href
  {https://ui.adsabs.harvard.edu/abs/1988SJNA...25..294E} {25, 294}

\bibitem[\protect\citeauthoryear{{Filippova}, {Mereminskiy}, {Lutovinov},
  {Molkov}  \& {Tsygankov}}{{Filippova} et~al.}{2017}]{Filippova+2017}
{Filippova} E.~V.,  {Mereminskiy} I.~A.,  {Lutovinov} A.~A.,  {Molkov} S.~V.,
  {Tsygankov} S.~S.,  2017, \mn@doi [Astronomy Letters]
  {10.1134/S1063773717110020}, \href
  {https://ui.adsabs.harvard.edu/abs/2017AstL...43..706F} {43, 706}

\bibitem[\protect\citeauthoryear{{Fukue}}{{Fukue}}{2019}]{fukue19}
{Fukue} J.,  2019, \mn@doi [\mnras] {10.1093/mnras/sty3286}, \href
  {https://ui.adsabs.harvard.edu/abs/2019MNRAS.483.2538F} {483, 2538}

\bibitem[\protect\citeauthoryear{{F{\"u}rst} et~al.,}{{F{\"u}rst}
  et~al.}{2016}]{fuerst16}
{F{\"u}rst} F.,  et~al., 2016, \mn@doi [\apjl] {10.3847/2041-8205/831/2/L14},
  \href {https://ui.adsabs.harvard.edu/abs/2016ApJ...831L..14F} {831, L14}

\bibitem[\protect\citeauthoryear{Harten, Lax  \& van Leer}{Harten
  et~al.}{1983}]{HLL}
Harten A.,  Lax P.,   van Leer B.,  1983, SIAM Rev, 25, 35

\bibitem[\protect\citeauthoryear{{Inoue}}{{Inoue}}{1975}]{Inoue1975}
{Inoue} H.,  1975, \pasj, \href
  {https://ui.adsabs.harvard.edu/abs/1975PASJ...27..311I} {27, 311}

\bibitem[\protect\citeauthoryear{{Israel} et~al.,}{{Israel}
  et~al.}{2017a}]{israel17a}
{Israel} G.~L.,  et~al., 2017a, \mn@doi [Science] {10.1126/science.aai8635},
  \href {https://ui.adsabs.harvard.edu/abs/2017Sci...355..817I} {355, 817}

\bibitem[\protect\citeauthoryear{{Israel} et~al.,}{{Israel}
  et~al.}{2017b}]{israel17b}
{Israel} G.~L.,  et~al., 2017b, \mn@doi [\mnras] {10.1093/mnrasl/slw218}, \href
  {https://ui.adsabs.harvard.edu/abs/2017MNRAS.466L..48I} {466, L48}

\bibitem[\protect\citeauthoryear{{Jahoda}, {Markwardt}, {Radeva}, {Rots},
  {Stark}, {Swank}, {Strohmayer}  \& {Zhang}}{{Jahoda} et~al.}{2006}]{jahoda06}
{Jahoda} K.,  {Markwardt} C.~B.,  {Radeva} Y.,  {Rots} A.~H.,  {Stark} M.~J.,
  {Swank} J.~H.,  {Strohmayer} T.~E.,   {Zhang} W.,  2006, \mn@doi [\apjs]
  {10.1086/500659}, \href
  {https://ui.adsabs.harvard.edu/abs/2006ApJS..163..401J} {163, 401}

\bibitem[\protect\citeauthoryear{{Jernigan}, {Klein}  \& {Arons}}{{Jernigan}
  et~al.}{2000}]{jernigan00}
{Jernigan} J.~G.,  {Klein} R.~I.,   {Arons} J.,  2000, \mn@doi [\apj]
  {10.1086/308390}, \href
  {https://ui.adsabs.harvard.edu/abs/2000ApJ...530..875J} {530, 875}

\bibitem[\protect\citeauthoryear{{Kaaret}, {Feng}  \& {Roberts}}{{Kaaret}
  et~al.}{2017}]{kaaret}
{Kaaret} P.,  {Feng} H.,   {Roberts} T.~P.,  2017, \mn@doi [\araa]
  {10.1146/annurev-astro-091916-055259}, \href
  {https://ui.adsabs.harvard.edu/abs/2017ARA&A..55..303K} {55, 303}

\bibitem[\protect\citeauthoryear{{Kawashima} \& {Ohsuga}}{{Kawashima} \&
  {Ohsuga}}{2020}]{Kawashima-Ohsuga2020}
{Kawashima} T.,  {Ohsuga} K.,  2020, \mn@doi [\pasj] {10.1093/pasj/psz136},
  \href {https://ui.adsabs.harvard.edu/abs/2020PASJ...72...15K} {72, 15}

\bibitem[\protect\citeauthoryear{{Kawashima}, {Mineshige}, {Ohsuga}  \&
  {Ogawa}}{{Kawashima} et~al.}{2016}]{Kawashima+2016}
{Kawashima} T.,  {Mineshige} S.,  {Ohsuga} K.,   {Ogawa} T.,  2016, \mn@doi
  [\pasj] {10.1093/pasj/psw075}, \href
  {https://ui.adsabs.harvard.edu/abs/2016PASJ...68...83K} {68, 83}

\bibitem[\protect\citeauthoryear{Keshtiban, Belblidia  \& Webster}{Keshtiban
  et~al.}{2003}]{lowmach}
Keshtiban I.,  Belblidia F.,   Webster M.,  2003, Int. J. Numer. Methods
  Fluids, 23, 77

\bibitem[\protect\citeauthoryear{{Klein} \& {Arons}}{{Klein} \&
  {Arons}}{1989}]{klein-arons1989}
{Klein} R.~I.,  {Arons} J.,  1989, in {Hunt} J.,  {Battrick} B.,  eds,  ESA
  Special Publication Vol. 1, Two Topics in X-Ray Astronomy, Volume 1: X Ray
  Binaries. Volume 2: AGN and the X Ray Background. p.~89

\bibitem[\protect\citeauthoryear{{Klein}, {Jernigan}, {Arons}, {Morgan}  \&
  {Zhang}}{{Klein} et~al.}{1996}]{klein96}
{Klein} R.~I.,  {Jernigan} J.~G.,  {Arons} J.,  {Morgan} E.~H.,   {Zhang} W.,
  1996, \mn@doi [\apjl] {10.1086/310277}, \href
  {http://adsabs.harvard.edu/abs/1996ApJ...469L.119K} {469, L119}

\bibitem[\protect\citeauthoryear{{Koldoba}, {Ustyugova}, {Romanova}  \&
  {Lovelace}}{{Koldoba} et~al.}{2008}]{koldoba08}
{Koldoba} A.~V.,  {Ustyugova} G.~V.,  {Romanova} M.~M.,   {Lovelace} R.~V.~E.,
  2008, \mn@doi [\mnras] {10.1111/j.1365-2966.2008.13394.x}, \href
  {https://ui.adsabs.harvard.edu/abs/2008MNRAS.388..357K} {388, 357}

\bibitem[\protect\citeauthoryear{{Kosec}, {Pinto}, {Walton}, {Fabian},
  {Bachetti}, {Brightman}, {F{\"u}rst}  \& {Grefenstette}}{{Kosec}
  et~al.}{2018}]{Kosec+2018}
{Kosec} P.,  {Pinto} C.,  {Walton} D.~J.,  {Fabian} A.~C.,  {Bachetti} M.,
  {Brightman} M.,  {F{\"u}rst} F.,   {Grefenstette} B.~W.,  2018, \mn@doi
  [\mnras] {10.1093/mnras/sty1626}, \href
  {https://ui.adsabs.harvard.edu/abs/2018MNRAS.479.3978K} {479, 3978}

\bibitem[\protect\citeauthoryear{{Kretschmar} et~al.,}{{Kretschmar}
  et~al.}{2021}]{velaX1}
{Kretschmar} P.,  et~al., 2021, \mn@doi [\aap] {10.1051/0004-6361/202040272},
  \href {https://ui.adsabs.harvard.edu/abs/2021A&A...652A..95K} {652, A95}

\bibitem[\protect\citeauthoryear{{Kulkarni} \& {Romanova}}{{Kulkarni} \&
  {Romanova}}{2013}]{Kulkarni-Romanova-2013}
{Kulkarni} A.~K.,  {Romanova} M.~M.,  2013, \mn@doi [\mnras]
  {10.1093/mnras/stt945}, \href
  {https://ui.adsabs.harvard.edu/abs/2013MNRAS.433.3048K} {433, 3048}

\bibitem[\protect\citeauthoryear{{Kulsrud} \& {Sunyaev}}{{Kulsrud} \&
  {Sunyaev}}{2020}]{Kulsrud-Sunyaev}
{Kulsrud} R.~M.,  {Sunyaev} R.,  2020, \mn@doi [Journal of Plasma Physics]
  {10.1017/S0022377820001026}, \href
  {https://ui.adsabs.harvard.edu/abs/2020JPlPh..86f9002K} {86, 905860602}

\bibitem[\protect\citeauthoryear{{Lamb}, {Pethick}  \& {Pines}}{{Lamb}
  et~al.}{1973}]{Lamb_etal1973}
{Lamb} F.~K.,  {Pethick} C.~J.,   {Pines} D.,  1973, \mn@doi [\apj]
  {10.1086/152325}, \href {http://adsabs.harvard.edu/abs/1973ApJ...184..271L}
  {184, 271}

\bibitem[\protect\citeauthoryear{{Levinson} \& {Nakar}}{{Levinson} \&
  {Nakar}}{2020}]{LN20}
{Levinson} A.,  {Nakar} E.,  2020, \mn@doi [\physrep]
  {10.1016/j.physrep.2020.04.003}, \href
  {https://ui.adsabs.harvard.edu/abs/2020PhR...866....1L} {866, 1}

\bibitem[\protect\citeauthoryear{{Li}}{{Li}}{2007}]{Li+2007}
{Li} T.-P.,  2007, \mn@doi [Nuclear Physics B Proceedings Supplements]
  {10.1016/j.nuclphysbps.2006.12.070}, \href
  {https://ui.adsabs.harvard.edu/abs/2007NuPhS.166..131L} {166, 131}

\bibitem[\protect\citeauthoryear{{Lipunova}, {Malanchev}, {Tsygankov},
  {Shakura}, {Tavleev}  \& {Kolesnikov}}{{Lipunova} et~al.}{2022}]{lipunova22}
{Lipunova} G.,  {Malanchev} K.,  {Tsygankov} S.,  {Shakura} N.,  {Tavleev} A.,
   {Kolesnikov} D.,  2022, \mn@doi [\mnras] {10.1093/mnras/stab3343}, \href
  {https://ui.adsabs.harvard.edu/abs/2022MNRAS.510.1837L} {510, 1837}

\bibitem[\protect\citeauthoryear{{Litwin}, {Brown}  \& {Rosner}}{{Litwin}
  et~al.}{2001}]{litwin01}
{Litwin} C.,  {Brown} E.~F.,   {Rosner} R.,  2001, \mn@doi [\apj]
  {10.1086/320952}, \href
  {https://ui.adsabs.harvard.edu/abs/2001ApJ...553..788L} {553, 788}

\bibitem[\protect\citeauthoryear{{Lyubarskii} \& {Syunyaev}}{{Lyubarskii} \&
  {Syunyaev}}{1988}]{Lyubarskii-Syunyaev1988}
{Lyubarskii} Y.~E.,  {Syunyaev} R.~A.,  1988, Soviet Astronomy Letters, \href
  {https://ui.adsabs.harvard.edu/abs/1988SvAL...14..390L} {14, 390}

\bibitem[\protect\citeauthoryear{{Meisel}, {Deibel}, {Keek}, {Shternin}  \&
  {Elfritz}}{{Meisel} et~al.}{2018}]{meisel18}
{Meisel} Z.,  {Deibel} A.,  {Keek} L.,  {Shternin} P.,   {Elfritz} J.,  2018,
  \mn@doi [Journal of Physics G Nuclear Physics] {10.1088/1361-6471/aad171},
  \href {https://ui.adsabs.harvard.edu/abs/2018JPhG...45i3001M} {45, 093001}

\bibitem[\protect\citeauthoryear{{Melatos} \& {Phinney}}{{Melatos} \&
  {Phinney}}{2001}]{MP01}
{Melatos} A.,  {Phinney} E.~S.,  2001, \mn@doi [\pasa] {10.1071/AS01056}, \href
  {https://ui.adsabs.harvard.edu/abs/2001PASA...18..421M} {18, 421}

\bibitem[\protect\citeauthoryear{{Miyamoto}, {Kimura}, {Kitamoto}, {Dotani}  \&
  {Ebisawa}}{{Miyamoto} et~al.}{1991}]{miyamoto91}
{Miyamoto} S.,  {Kimura} K.,  {Kitamoto} S.,  {Dotani} T.,   {Ebisawa} K.,
  1991, \mn@doi [\apj] {10.1086/170837}, \href
  {https://ui.adsabs.harvard.edu/abs/1991ApJ...383..784M} {383, 784}

\bibitem[\protect\citeauthoryear{{Mukherjee}, {Bhattacharya}  \&
  {Mignone}}{{Mukherjee} et~al.}{2013}]{mukherjee13}
{Mukherjee} D.,  {Bhattacharya} D.,   {Mignone} A.,  2013, \mn@doi [\mnras]
  {10.1093/mnras/stt1344}, \href
  {https://ui.adsabs.harvard.edu/abs/2013MNRAS.435..718M} {435, 718}

\bibitem[\protect\citeauthoryear{{Mushtukov} \& {Tsygankov}}{{Mushtukov} \&
  {Tsygankov}}{2022}]{Mushtukov-Tsygankov2022}
{Mushtukov} A.,  {Tsygankov} S.,  2022, arXiv e-prints, \href
  {https://ui.adsabs.harvard.edu/abs/2022arXiv220414185M} {p. arXiv:2204.14185}

\bibitem[\protect\citeauthoryear{{Mushtukov}, {Suleimanov}, {Tsygankov}  \&
  {Ingram}}{{Mushtukov} et~al.}{2017}]{mushtukov17}
{Mushtukov} A.~A.,  {Suleimanov} V.~F.,  {Tsygankov} S.~S.,   {Ingram} A.,
  2017, \mn@doi [\mnras] {10.1093/mnras/stx141}, \href
  {https://ui.adsabs.harvard.edu/abs/2017MNRAS.467.1202M} {467, 1202}

\bibitem[\protect\citeauthoryear{{Mushtukov}, {Portegies Zwart}, {Tsygankov},
  {Nagirner}  \& {Poutanen}}{{Mushtukov} et~al.}{2021}]{Mushtukov+2021}
{Mushtukov} A.~A.,  {Portegies Zwart} S.,  {Tsygankov} S.~S.,  {Nagirner}
  D.~I.,   {Poutanen} J.,  2021, \mn@doi [\mnras] {10.1093/mnras/staa3809},
  \href {https://ui.adsabs.harvard.edu/abs/2021MNRAS.501.2424M} {501, 2424}

\bibitem[\protect\citeauthoryear{{Nowak}, {Vaughan}, {Wilms}, {Dove}  \&
  {Begelman}}{{Nowak} et~al.}{1999}]{nowak99}
{Nowak} M.~A.,  {Vaughan} B.~A.,  {Wilms} J.,  {Dove} J.~B.,   {Begelman}
  M.~C.,  1999, \mn@doi [\apj] {10.1086/306610}, \href
  {https://ui.adsabs.harvard.edu/abs/1999ApJ...510..874N} {510, 874}

\bibitem[\protect\citeauthoryear{{Parfrey} \& {Tchekhovskoy}}{{Parfrey} \&
  {Tchekhovskoy}}{2017}]{parfrey17}
{Parfrey} K.,  {Tchekhovskoy} A.,  2017, \mn@doi [\apjl]
  {10.3847/2041-8213/aa9c85}, \href
  {https://ui.adsabs.harvard.edu/abs/2017ApJ...851L..34P} {851, L34}

\bibitem[\protect\citeauthoryear{{Parfrey}, {Spitkovsky}  \&
  {Beloborodov}}{{Parfrey} et~al.}{2016}]{parfrey16}
{Parfrey} K.,  {Spitkovsky} A.,   {Beloborodov} A.~M.,  2016, \mn@doi [\apj]
  {10.3847/0004-637X/822/1/33}, \href
  {https://ui.adsabs.harvard.edu/abs/2016ApJ...822...33P} {822, 33}

\bibitem[\protect\citeauthoryear{{Patruno} \& {Watts}}{{Patruno} \&
  {Watts}}{2021}]{AMXP_review}
{Patruno} A.,  {Watts} A.~L.,  2021, in {Belloni} T.~M.,  {M{\'e}ndez} M.,
  {Zhang} C.,  eds,  Astrophysics and Space Science Library Vol. 461,
  Astrophysics and Space Science Library. pp 143--208 (\mn@eprint {arXiv}
  {1206.2727}), \mn@doi{10.1007/978-3-662-62110-3\_4}

\bibitem[\protect\citeauthoryear{{Phinney}}{{Phinney}}{1982}]{phinney82}
{Phinney} E.~S.,  1982, \mn@doi [\mnras] {10.1093/mnras/198.4.1109}, \href
  {https://ui.adsabs.harvard.edu/abs/1982MNRAS.198.1109P} {198, 1109}

\bibitem[\protect\citeauthoryear{{Pinto}, {Middleton}  \& {Fabian}}{{Pinto}
  et~al.}{2016}]{Pinto17}
{Pinto} C.,  {Middleton} M.~J.,   {Fabian} A.~C.,  2016, \mn@doi [\nat]
  {10.1038/nature17417}, \href
  {https://ui.adsabs.harvard.edu/abs/2016Natur.533...64P} {533, 64}

\bibitem[\protect\citeauthoryear{{Pinto} et~al.,}{{Pinto}
  et~al.}{2021}]{pinto21}
{Pinto} C.,  et~al., 2021, \mn@doi [\mnras] {10.1093/mnras/stab1648}, \href
  {https://ui.adsabs.harvard.edu/abs/2021MNRAS.505.5058P} {505, 5058}

\bibitem[\protect\citeauthoryear{{Pintore}, {Zampieri}, {Stella}, {Wolter},
  {Mereghetti}  \& {Israel}}{{Pintore} et~al.}{2017}]{pintore17}
{Pintore} F.,  {Zampieri} L.,  {Stella} L.,  {Wolter} A.,  {Mereghetti} S.,
  {Israel} G.~L.,  2017, \mn@doi [\apj] {10.3847/1538-4357/836/1/113}, \href
  {https://ui.adsabs.harvard.edu/abs/2017ApJ...836..113P} {836, 113}

\bibitem[\protect\citeauthoryear{{Portegies Zwart}, {Dewi}  \&
  {Maccarone}}{{Portegies Zwart} et~al.}{2004}]{PZ04}
{Portegies Zwart} S.~F.,  {Dewi} J.,   {Maccarone} T.,  2004, \mn@doi [\mnras]
  {10.1111/j.1365-2966.2004.08327.x}, \href
  {https://ui.adsabs.harvard.edu/abs/2004MNRAS.355..413P} {355, 413}

\bibitem[\protect\citeauthoryear{{Poutanen}, {Lipunova}, {Fabrika}, {Butkevich}
   \& {Abolmasov}}{{Poutanen} et~al.}{2007}]{poutanen07}
{Poutanen} J.,  {Lipunova} G.,  {Fabrika} S.,  {Butkevich} A.~G.,   {Abolmasov}
  P.,  2007, \mn@doi [\mnras] {10.1111/j.1365-2966.2007.11668.x}, \href
  {https://ui.adsabs.harvard.edu/abs/2007MNRAS.377.1187P} {377, 1187}

\bibitem[\protect\citeauthoryear{{Pringle} \& {Rees}}{{Pringle} \&
  {Rees}}{1972}]{pringle-rees1972}
{Pringle} J.~E.,  {Rees} M.~J.,  1972, \aap, \href
  {http://adsabs.harvard.edu/abs/1972A%26A....21....1P} {21, 1}

\bibitem[\protect\citeauthoryear{{Raguzova} \& {Popov}}{{Raguzova} \&
  {Popov}}{2005}]{raguzova}
{Raguzova} N.~V.,  {Popov} S.~B.,  2005, \mn@doi [Astronomical and
  Astrophysical Transactions] {10.1080/10556790500497311}, \href
  {https://ui.adsabs.harvard.edu/abs/2005A&AT...24..151R} {24, 151}

\bibitem[\protect\citeauthoryear{Ramachandran \& Varoquaux}{Ramachandran \&
  Varoquaux}{2011}]{mayavi}
Ramachandran P.,  Varoquaux G.,  2011, Computing in Science \& Engineering, 13,
  40

\bibitem[\protect\citeauthoryear{{Rappaport} \& {Joss}}{{Rappaport} \&
  {Joss}}{1983}]{RJ83}
{Rappaport} S.~A.,  {Joss} P.~C.,  1983, in {Lewin} W.~H.~G.,  {van den Heuvel}
  E.~P.~J.,  eds, Accretion-Driven Stellar X-ray Sources. pp 1--39

\bibitem[\protect\citeauthoryear{{Reig}}{{Reig}}{2011}]{reig11}
{Reig} P.,  2011, \mn@doi [\apss] {10.1007/s10509-010-0575-8}, \href
  {https://ui.adsabs.harvard.edu/abs/2011Ap&SS.332....1R} {332, 1}

\bibitem[\protect\citeauthoryear{{Reig} \& {Nespoli}}{{Reig} \&
  {Nespoli}}{2013}]{RN13}
{Reig} P.,  {Nespoli} E.,  2013, \mn@doi [\aap] {10.1051/0004-6361/201219806},
  \href {https://ui.adsabs.harvard.edu/abs/2013A&A...551A...1R} {551, A1}

\bibitem[\protect\citeauthoryear{{Revnivtsev}, {Churazov}, {Postnov}  \&
  {Tsygankov}}{{Revnivtsev} et~al.}{2009}]{revnivtsev09}
{Revnivtsev} M.,  {Churazov} E.,  {Postnov} K.,   {Tsygankov} S.,  2009,
  \mn@doi [\aap] {10.1051/0004-6361/200912317}, \href
  {http://adsabs.harvard.edu/abs/2009A%26A...507.1211R} {507, 1211}

\bibitem[\protect\citeauthoryear{{Revnivtsev}, {Molkov}  \&
  {Pavlinsky}}{{Revnivtsev} et~al.}{2015}]{revnivtsev15}
{Revnivtsev} M.~G.,  {Molkov} S.~V.,   {Pavlinsky} M.~N.,  2015, \mn@doi
  [\mnras] {10.1093/mnras/stv1263}, \href
  {https://ui.adsabs.harvard.edu/abs/2015MNRAS.451.4253R} {451, 4253}

\bibitem[\protect\citeauthoryear{{Rodr\'{i}guez Castillo}
  et~al.,}{{Rodr\'{i}guez Castillo} et~al.}{2020}]{castillo_m51}
{Rodr\'{i}guez Castillo} G.~A.,  et~al., 2020, \mn@doi [\apj]
  {10.3847/1538-4357/ab8a44}, \href
  {https://ui.adsabs.harvard.edu/abs/2020ApJ...895...60R} {895, 60}

\bibitem[\protect\citeauthoryear{{Romanova}, {Ustyugova}, {Koldoba}, {Wick}  \&
  {Lovelace}}{{Romanova} et~al.}{2003}]{romanova03}
{Romanova} M.~M.,  {Ustyugova} G.~V.,  {Koldoba} A.~V.,  {Wick} J.~V.,
  {Lovelace} R.~V.~E.,  2003, \mn@doi [\apj] {10.1086/377514}, \href
  {https://ui.adsabs.harvard.edu/abs/2003ApJ...595.1009R} {595, 1009}

\bibitem[\protect\citeauthoryear{{Santangelo}, {del Sordo}, {Segreto}, {dal
  Fiume}, {Orlandini}  \& {Piraino}}{{Santangelo} et~al.}{1998}]{santangelo98}
{Santangelo} A.,  {del Sordo} S.,  {Segreto} A.,  {dal Fiume} D.,  {Orlandini}
  M.,   {Piraino} S.,  1998, \aap, \href
  {https://ui.adsabs.harvard.edu/abs/1998A&A...340L..55S} {340, L55}

\bibitem[\protect\citeauthoryear{{Scharlemann}}{{Scharlemann}}{1978}]{scharlemann78}
{Scharlemann} E.~T.,  1978, \mn@doi [\apj] {10.1086/155823}, \href
  {https://ui.adsabs.harvard.edu/abs/1978ApJ...219..617S} {219, 617}

\bibitem[\protect\citeauthoryear{Tikhonov \& Samarskii}{Tikhonov \&
  Samarskii}{2013}]{mathphys}
Tikhonov A.,  Samarskii A.,  2013, Equations of Mathematical Physics.
Dover Books on Physics, Dover Publications, \url
  {https://books.google.fi/books?id=PTmoAAAAQBAJ}

\bibitem[\protect\citeauthoryear{{Tolstov}, {Blinnikov}, {Nagataki}  \&
  {Nomoto}}{{Tolstov} et~al.}{2015}]{tolstov15}
{Tolstov} A.,  {Blinnikov} S.,  {Nagataki} S.,   {Nomoto} K.,  2015, \mn@doi
  [\apj] {10.1088/0004-637X/811/1/47}, \href
  {https://ui.adsabs.harvard.edu/abs/2015ApJ...811...47T} {811, 47}

\bibitem[\protect\citeauthoryear{{Toro}, {Spruce}  \& {Speares}}{{Toro}
  et~al.}{1994}]{toro94}
{Toro} E.~F.,  {Spruce} M.,   {Speares} W.,  1994, \mn@doi [Shock Waves]
  {10.1007/BF01414629}, \href
  {http://adsabs.harvard.edu/abs/1994ShWav...4...25T} {4, 25}

\bibitem[\protect\citeauthoryear{{Tsygankov}, {Lutovinov}, {Doroshenko},
  {Mushtukov}, {Suleimanov}  \& {Poutanen}}{{Tsygankov}
  et~al.}{2016}]{Tsygankov16}
{Tsygankov} S.~S.,  {Lutovinov} A.~A.,  {Doroshenko} V.,  {Mushtukov} A.~A.,
  {Suleimanov} V.,   {Poutanen} J.,  2016, \mn@doi [\aap]
  {10.1051/0004-6361/201628236}, \href
  {https://ui.adsabs.harvard.edu/abs/2016A&A...593A..16T} {593, A16}

\bibitem[\protect\citeauthoryear{{Turkel}}{{Turkel}}{1999}]{turkel99}
{Turkel} E.,  1999, \mn@doi [Annual Review of Fluid Mechanics]
  {10.1146/annurev.fluid.31.1.385}, \href
  {https://ui.adsabs.harvard.edu/abs/1999AnRFM..31..385T} {31, 385}

\bibitem[\protect\citeauthoryear{{West}, {Wolfram}  \& {Becker}}{{West}
  et~al.}{2017a}]{West+2017}
{West} B.~F.,  {Wolfram} K.~D.,   {Becker} P.~A.,  2017a, \mn@doi [\apj]
  {10.3847/1538-4357/835/2/129}, \href
  {https://ui.adsabs.harvard.edu/abs/2017ApJ...835..129W} {835, 129}

\bibitem[\protect\citeauthoryear{{West}, {Wolfram}  \& {Becker}}{{West}
  et~al.}{2017b}]{West17b}
{West} B.~F.,  {Wolfram} K.~D.,   {Becker} P.~A.,  2017b, \mn@doi [\apj]
  {10.3847/1538-4357/835/2/130}, \href
  {https://ui.adsabs.harvard.edu/abs/2017ApJ...835..130W} {835, 130}

\bibitem[\protect\citeauthoryear{{Wijnands} \& {van der Klis}}{{Wijnands} \&
  {van der Klis}}{1999}]{WK99}
{Wijnands} R.,  {van der Klis} M.,  1999, \mn@doi [\apj] {10.1086/306993},
  \href {https://ui.adsabs.harvard.edu/abs/1999ApJ...514..939W} {514, 939}

\bibitem[\protect\citeauthoryear{{Wolff} et~al.,}{{Wolff}
  et~al.}{2019}]{wolff_baas}
{Wolff} M.,  et~al., 2019, \baas, \href
  {https://ui.adsabs.harvard.edu/abs/2019BAAS...51c.386W} {51, 386}

\bibitem[\protect\citeauthoryear{{Zel'dovich} \& {Raizer}}{{Zel'dovich} \&
  {Raizer}}{1967}]{zeldovich-raizer}
{Zel'dovich} Y.~B.,  {Raizer} Y.~P.,  1967, {Physics of shock waves and
  high-temperature hydrodynamic phenomena}

\bibitem[\protect\citeauthoryear{{Zhang}, {Blaes}  \& {Jiang}}{{Zhang}
  et~al.}{2021}]{Zhang+2021}
{Zhang} L.,  {Blaes} O.,   {Jiang} Y.-F.,  2021, \mn@doi [\mnras]
  {10.1093/mnras/stab2510}, \href
  {https://ui.adsabs.harvard.edu/abs/2021MNRAS.508..617Z} {508, 617}

\bibitem[\protect\citeauthoryear{{Zhang}, {Blaes}  \& {Jiang}}{{Zhang}
  et~al.}{2022}]{zhang22}
{Zhang} L.,  {Blaes} O.,   {Jiang} Y.-F.,  2022, \mn@doi [\mnras]
  {10.1093/mnras/stac1815}, \href
  {https://ui.adsabs.harvard.edu/abs/2022MNRAS.515.4371Z} {515, 4371}

\bibitem[\protect\citeauthoryear{{Zhang}, {Blaes}  \& {Jiang}}{{Zhang}
  et~al.}{2023}]{zhang23}
{Zhang} L.,  {Blaes} O.,   {Jiang} Y.-F.,  2023, \mn@doi [\mnras]
  {10.1093/mnras/stad063}, \href
  {https://ui.adsabs.harvard.edu/abs/2023MNRAS.520.1421Z} {520, 1421}

\bibitem[\protect\citeauthoryear{{van Straaten}, {van der Klis}  \&
  {Wijnands}}{{van Straaten} et~al.}{2005}]{vstraaten05}
{van Straaten} S.,  {van der Klis} M.,   {Wijnands} R.,  2005, \mn@doi [\apj]
  {10.1086/426183}, \href
  {https://ui.adsabs.harvard.edu/abs/2005ApJ...619..455V} {619, 455}

\bibitem[\protect\citeauthoryear{{van der Klis}}{{van der
  Klis}}{2000}]{vdK_review}
{van der Klis} M.,  2000, \mn@doi [\araa] {10.1146/annurev.astro.38.1.717},
  \href {https://ui.adsabs.harvard.edu/abs/2000ARA&A..38..717V} {38, 717}

\makeatother
\end{thebibliography}




\appendix

\section{Geometry: technical details}\label{app:geo} 
\def\R0{R_{\rm e}}

The field of a magnetic dipole is given by the general formula
\begin{equation}\label{eq:Bdipole}
 \vector{B} = \frac{3(\vector{\mu} \cdot \vector{R}) \vector{R}}{R^5} - \frac{\vector{\mu}}{R^3},
 \end{equation}
where $\vector{\mu}$ is the magnetic dipole moment vector. 
We assume the dipole to be aligned with the angular momentum in the accretion disc (considered only as a source of mass) and with the rotation of the NS. 
We also assume that the field lines retain their shapes (there is no significant back-reaction from the flow), and the matter follows the field lines. 
The geometry of the flow in these assumptions implies a simple relation valid for an aligned dipole in spherical coordinates $R=\R0\sin^2 \theta$, where $\theta$ is the polar angle, and $R$ is the spherical radial coordinate. 

A unit vector along the field line
\begin{equation}
\vector{e}_l = \frac{2\cos\theta\, \vector{e}_R + \sin\theta\, \vector{e}_\theta}{\sqrt{3\cos^2\theta+1}},
\end{equation}
where $\vector{e}_R$ and $\vector{e}_\theta$ are the corresponding unit coordinate vectors. 
The cosine of the angle between the magnetic field and the radial unit vector is the scalar product of $\vector{e}_r$ and $\vector{e}_l$. 
Taking the scalar product allows us to relate $R$ and the coordinate along the field line $l$ as
\begin{equation}\label{E:cosa}
\frac{{\rm d} R}{{\rm d} l} = \frac{2\cos \theta}{\sqrt{3\cos^2\theta +1}} .
\end{equation}
The size of the flow perpendicular to the field line $\delta \ll R$ (width of the accretion curtain) may be calculated using flux conservation. 
Magnetic flux $|\vector{B}| \,A_\perp = const  $, where $|\vector{B}|  = \mu \sqrt{3\, \cos^2\theta +1} / R^3$ follows  from equation~\eqref{eq:Bdipole}. 
The cross-section of the flux tube equals
\begin{equation}
    A_\perp = 4\pi\, a\, R\sin \theta\, \delta,
\end{equation}
which implies
\begin{equation}\label{E:dcross}
\frac{\delta}{R}  = \frac{\sin\theta}{\sqrt{1+3\cos^2 \theta}} \, \frac{ \Delta
  \R0}{\R0}.
\end{equation}

The polar angle of the footpoint of a field line is given by the relation
\begin{equation}
\label{E:sintheta_min}
\sin\theta_{\rm min} = \sqrt{R_\NS/R_{\rm e}}    . 
\end{equation}
Substituting this expression to (\ref{E:dcross}), we find the width of the accretion curtain at the surface
\begin{equation}\label{eq:app:deltastar}
\delta_\NS = R_\NS \, \sqrt{\frac{R_\NS}{R_{\rm e} (4-3 R_\NS/R_{\rm e}) } }    \frac{ \Delta
  \R0}{\R0} \approx R_\NS \,
\frac{\Delta   \R0}{2\, \R0}\,\sqrt{\frac{R_\NS}{R_{\rm e}} }\, .
\end{equation}
The part of the NS surface occupied by the accretion flows is
\begin{equation}
\label{eq.Aperp}
\frac{A_{\perp,\NS}}{4\uppi R_\NS^2} = \frac{a\,\delta_\NS}{\sqrt{R_\NS R_{\rm e}}} \approx
\frac {a}{2} \, \frac{ R_\NS}{R_{\rm e}}\, \frac{\Delta   \R0}{R_{\rm e}} .
\end{equation}
The approximations above hold if $R_\NS \ll \R0$ and $\Delta \R0 \ll \R0$. 
If the size of the magnetosphere is two orders of magnitude larger than the radius of the NS, $A_{\perp, \NS}$ is expected to be smaller than  one per cent of the surface  of the star.

The angle $\alpha$ between the magnetic line and horizontal direction  may be found by considering the unit vector
\begin{equation}
    \vector{e}_\varpi = \vector{e}_R \sin\theta + \vector{e}_\theta \cos\theta,
\end{equation}
directed along the radial direction of a cylindrical coordinate system. Scalar product $\vector{e}_\varpi \cdot \vector{e}_l$ allows one to calculate the angle $\alpha$ between the cylindric radial direction set by vector $\vector{e}_\varpi$ and the field line (see Fig.~\ref{fig:nssketch}), 
\begin{equation}
\cos \alpha = \frac{3\cos\theta\, \sin\theta}{\sqrt{1+3\cos^2\theta}}    \, .
\end{equation}
The sign of $\alpha$ is assumed positive if the tangent to the field line makes an acute angle with the symmetry axis (as in Fig.~\ref{fig:nssketch}). 
Note that $\alpha$ changes sign when $\cos \theta = 1/\sqrt{3}$. 
Centrifugal force, projected onto the field line, enters the momentum conservation equation with a multiplier of $\cos\alpha$. 
Gravity is directed along $\vector{e}_R$, hence its contribution to equation~(\ref{E:src:s}) should be multiplied by
\begin{equation}
    \vector{e}_R \cdot \vector{e}_l = \sin(\alpha +\theta) = \frac{2\cos\theta}{\sqrt{1+3\cos^2\theta}} .
\end{equation}

\section{Gas-to-total pressure ratio}\label{app:beta}

Pressure ratio $\beta$ is defined as 
\begin{equation}
\beta = p_{\rm gas}/p = \frac{p_{\rm gas}}{ p_{\rm gas} + p_{\rm rad}},
\end{equation}
where $p_{\rm gas}$ is gas pressure
\begin{equation}
    p_{\rm gas} = \frac{\rho kT}{\tilde{m}},
\end{equation}
$\tilde{m}$ is the mean mass of a particle, and
\begin{equation}\label{E:app:prad}
    p_{\rm rad} = 
    \frac{u_{\rm rad}}{3}=
    \frac{4}{3} \frac{\sigma_{\rm SB} }{c} T^4.
\end{equation}
For the gas constituent of pressure, 
\begin{equation}\label{E:app:ugas}
u_{\rm gas} = \frac32 \, p_{\rm gas}  = \frac32 \frac{\rho}{\tilde{m}} kT\, .
\end{equation}

From equation~\eqref{E:app:prad} we express the temperature as
\begin{equation}
T = \left( \frac{c\,u_{\rm rad}}{4\,\sigma_{\rm SB} } \right)^{1/4} =  \left(  \frac{c\, u}{4\,\sigma_{\rm SB} } \frac{1-\beta}{1-\beta/2} \right)^{1/4},
\end{equation}
and substitute it to (\ref{E:app:ugas}) obtaining an equation for $\beta$, which may be solved given mass and energy density
\begin{equation}
\frac{\beta}{\left(1-\beta/2\right)^{3/4}\left( 1-\beta\right)^{1/4}}  = \frac{3}{\sqrt{2}} \frac{k}{\tilde{m}} \left(\frac{c}{\sigma_{\rm SB}} \right)^{1/4} \frac{\rho}{u^{3/4}}.
\end{equation}

\bsp	
\label{lastpage}
\end{document}